\documentclass[showpacs,twocolumn,floatfix,longbibliography,%
pra,aps]{revtex4-1}

\usepackage{amsmath}
\usepackage{amssymb}
\usepackage{maybemath}
\usepackage{xcolor}
\usepackage{graphicx}
\usepackage{bm}
\usepackage{diagbox}
\usepackage{siunitx}

\usepackage{placeins}

\usepackage{dcolumn}
\newcolumntype{.}{D{x}{.}{-1}}

\usepackage[%
colorlinks=true,
urlcolor=blue,
linkcolor=blue,
citecolor=blue
]{hyperref}

\definecolor{kellygreen}{rgb}{0.3, 0.73, 0.09}

\definecolor{garrosgreen}{rgb}{0.1, 0.4, 0.1}
\definecolor{dartmouthgreen}{rgb}{0.05, 0.5, 0.06}

\definecolor{duelferred}{rgb}{0.7, 0.2, 0.1}
\definecolor{cambridgeblue}{rgb}{0.1, 0.3, 1.0}
\definecolor{oxfordblue}{rgb}{0.05, 0.2, 0.7}

\definecolor{gold}{rgb}{0.85,.66,0}

\def\dd{{\mathrm{d}}}
\def\ii{{\mathrm{i}}}

\def\vdw{van der Waals}
\def\VDW{Van der Waals}
\def\cp{Casimir--Polder}

\def\calL{{\mathcal L}}
\def\calF{{\mathcal F}}
\def\calP{{\mathcal P}}
\def\calQ{{\mathcal Q}}

\def\calW{{\mathcal W}}

\newcommand{\dir}{{(\rm{dir})}}
\newcommand{\mix}{{(\rm{mix})}}

\begin{document}

\newcommand{\addrROLLA}{Department of Physics,
Missouri University of Science and Technology,
Rolla, Missouri 65409-0640, USA}

\newcommand{\addrHDphiltheo}{Institut f\"ur Theoretische Physik,
Universit\"{a}t Heidelberg,
Philosophenweg 16, 69120 Heidelberg, Germany}

\newcommand{\addrLEBEDEV}{P. N. Lebedev Physics
Institute, Leninsky prosp.~53, Moscow, 119991 Russia}

\newcommand{\addrMUC}{Max--Planck--Institut f\"ur 
Quantenoptik, Hans--Kopfermann-Stra\ss{}e~1, 
85748 Garching, Germany}

\newcommand{\addrBUDKER}{Budker Institute of Nuclear Physics, 
630090 Novosibirsk, Russia}

\newcommand{\addrRQC}{Russian Quantum Center,
Business-center ``Ural'', 100A Novaya street,
Skolkovo, Moscow, 143025 Russia}

\title{Long-range interactions of hydrogen atoms in excited states.~III.\\
\texorpdfstring{$\maybebm{nS}$}{nS}--\texorpdfstring{$\maybebm{1S}$}{1S} interactions
for \texorpdfstring{$\maybebm{n \geq 3}$}{n >= 3}}

\author{C. M. Adhikari}
\affiliation{\addrROLLA}

\author{V. Debierre}
\affiliation{\addrROLLA}

\author{U. D. Jentschura}
\affiliation{\addrROLLA}

\begin{abstract}
The long-range interaction of excited neutral atoms 
has a number of interesting and surprising properties, such 
as the prevalence of long-range, oscillatory 
tails, and the emergence of numerically 
large \vdw{} $C_6$ coefficients.
Furthermore, the energetically quasi-degenerate 
$nP$ states require special attention and lead 
to mathematical subtleties.
Here, we analyze the interaction of excited 
hydrogen atoms in $nS$ states ($3 \leq n \leq 12$)
with ground-state hydrogen atoms, 
and find that the $C_6$ coefficients 
roughly grow with the fourth power of the 
principal quantum number, and can 
reach values in excess of $240\,000$ (in atomic units)
for states with $n = 12$.
The nonretarded \vdw{} result is relevant 
to the distance range $R \ll a_0/\alpha$,
where $a_0$ is the Bohr radius and 
$\alpha$ is the fine-structure constant.
The \cp{} range encompasses the 
interatomic distance range $a_0/\alpha \ll R \ll \hbar c/\calL$,
where $\calL$ is the Lamb shift energy.
In this range, the contribution of quasi-degenerate excited $nP$ 
states remains nonretarded and competes with 
the $1/R^2$ and $1/R^4$ tails of the pole terms which 
are generated by lower-lying $mP$ states
with $2 \leq m \leq n-1$, due to virtual resonant emission.
The dominant pole terms are also analyzed in the 
Lamb shift range $R \gg \hbar c/\calL$.
The familiar $1/R^7$ asymptotics from the usual \cp{}
theory is found to be completely irrelevant for the 
analysis of excited-state interactions.
The calculations are carried out 
to high precision using computer algebra in order to 
handle a large number of terms in intermediate steps of the 
calculation, for highly excited states.
\end{abstract}

\pacs{31.30.jh, 31.30.J-, 31.30.jf}

\maketitle

%% \tableofcontents
%% \newpage

%
% Introduction
%
\section{Introduction}
\label{sec1}

In general, the analysis of long-range interactions among 
neutral atoms in excited states is less trivial 
than one would expect at first glance.  This is
true for hydrogen atoms (in excited $S$ states), which form the basis of the
current investigation, as much as any other atom.  The reasons are threefold.
First, we note the presence of quasi-degenerate excited $nP$ states, which are
only displaced from the $nS$ states by the Lamb shift or the fine 
structure~\cite{AdEtAl2017vdWi}.  Due to
the long wavelength of the involved virtual transitions, the contribution of
the quasi-degenerate states remains non-retarded over wide distance ranges.
Second, the presence of lower-lying virtual $mP$ states with $m \leq n$ 
leads to both oscillatory energy shifts and distance-dependent corrections to
the decay width of the excited state~\cite{SaKa2015,DoGuLa2015,JeAdDe2017prl}.  
Third, for $nS$--$1S$ interactions,
there is a {\em gerade}--{\em ungerade} mixing term that depends on the
symmetry of the excited-state contributions to the two-atom wave function.  The
mixing term is numerically large for 
$2S$--$1S$ interactions~\cite{Ch1972,AdEtAl2017vdWi}.
The eigenstates
of the total Hamiltonian are composed of coherent superpositions of $nS$--$1S$
and $1S$--$nS$ states in the two-atom system.

Let us try to provide some background on these issues.  We have recently
analyzed~\cite{AdEtAl2017vdWi} the interaction of metastable $2S$ hydrogen atoms with
ground-state atoms.
A long-standing discrepancy regarding the
numerical value of the \vdw{} $C_6$ coefficient could be resolved, 
and the mixing term was treated for
$2S$--$1S$ interactions~\cite{Ch1972,DeYo1973,TaCh1986,AdEtAl2017vdWi}. 
In~\cite{JeEtAl2017vdWii},
we have analyzed $2S$--$2S$ interactions, and we have determined the
hyperfine-resolved eigenstates of the \vdw{} interaction, both among the
$S$--$S$, $P$--$P$ as well as the $S$--$P$ submanifolds of the $n=2$ hydrogen
states. The physically interesting oscillatory tails of \vdw{} interactions
involving excited states were recently discussed in 
Refs.~\cite{DoGuLa2015,SaKa2015,JeDe2017vdw0}. The special role of
quasi-degenerate excited states was analyzed in~\cite{AdEtAl2017vdWi}. 
All of these concepts are relevant to the current investigation.

Finally, we should 
mention that the numerical evaluation of the \vdw{} $C_6$ coefficient 
for excited states demands the rather
sophisticated use of recurrence relations in order to express the
polarizability matrix elements in terms of hypergeometric functions.
This phenomenon is familiar from Lamb shift 
calculations~\cite{JePa1996,CzJePa2005}. The
numerical calculations lead to \vdw{} $C_6$ coefficients that grow rapidly
with the principal quantum number.

Throughout this article, we work in SI mksA units and keep all factors of $\hbar$
and $c$ in the formulas.  With this choice, we attempt to extend the
accessibility of the presentation to two different communities, namely, the
quantum electrodynamics community which in general uses the natural unit
system, and the atomic physics community where the atomic unit system is
canonically employed.  In the former, one sets $\hbar = c = \epsilon_0 = 1$,
the electron mass is denoted by $m_e$, and one has the relation $e^2 = 4\pi\alpha$.  
This unit system is used, e.g., in the investigation reported in
Ref.~\cite{Pa2005longrange} on relativistic corrections to the Casimir--Polder
interaction.  In the atomic unit system, one
has $\left|e\right| = \hbar = m_e = 1$, and $4 \pi \epsilon_0 = 1$.  The speed
of light, in the atomic unit system, is $c = 1/\alpha \approx 137.036$. This
system of units is especially useful for the analysis of atomic
properties without radiative corrections.  As the subject of the current study lies
in between the two mentioned fields of interest, we choose the SI MKSA unit system
as the most appropriate reference frame for our calculations. The formulas do
not become unnecessarily complex, and can be evaluated with ease for any
experimental application.

We organize this paper as follows. The problem is somewhat involved;
after an orientation (in Sec.~\ref{sec2}),
we focus on the $3S$--$1S$ interaction in
Sec.~\ref{sec3}.  In Sec.~\ref{subsec:CloseD}, we study the 
\vdw{} range. The very-large-distance limit is discussed in
Sec.~\ref{subsec:AwayD} (atomic distance larger than the wavelength of the Lamb
shift transition). The intermediate Casimir--Polder range 
is discussed in
Sec.~\ref{subsec:MidD}.  States with $4 \leq n \leq 12$ are
analyzed in Sec.~\ref{sec4}. Numerical examples are 
discussed in Sec.~\ref{sec5}. We summarize 
in Sec.~\ref{sec6}.

%
% Orientation
%
\section{Orientation}
\label{sec2}

As we are not interested in the hyperfine structure
of the excited $nS$ state ($n \geq 3$),
we may write the total Hamiltonian of the 
two-atom system as
\begin{equation} \label{eq:TotH}
H_{\rm total} = H_S + H_{\rm FS} + H_{\rm LS} + H_{\rm vdW} \,.
\end{equation}
Here, $H_S$ is the Schr\"{o}dinger Hamiltonian, while
$H_{\rm FS}$ is the fine structure Hamiltonian, 
which can be approximated as
(see Chap.~34 of Ref.~\cite{BeLiPi1982vol4})
\begin{align}
\label{eq:FineSH}
H_{\rm FS}= & \; 
\sum_{i=A,B} \left[ -\frac{\vec{p}_i^{\,4}}{8 m_e^3 c^2}
+ \frac{1}{2} \alpha\left(\frac{\hbar^2 g_s}{2 m_e^{2} \, c}\right)
\frac{\vec{L}_i\cdot\vec{S}_i}{| \vec r_i |^3}\right.
\nonumber\\
& \; \left.
+ \frac{\hbar^{3}}{8 m_e^{2} c}\,4\pi\alpha \, 
\delta^{(3)}(\vec{r}_i) \right],
\end{align}
where $m_e$ is the electron mass. 
The momenta of the two atomic electrons
are denoted by $\vec{p}_i$ 
(here, $i$ runs over the atoms $A$ and $B$),
and the distance vectors $\vec{r}_i = \vec x_i - \vec R_i$ are the 
coordinates relative to the nuclei
(the electron and nucleus coordinates 
are $\vec x_i$ and $\vec R_i$, respectively).
The fine-structure constant is denoted by
$\alpha \approx 1/137.036$, and the 
electronic $g$ factor is $g_s\simeq2.002\,319$.
As \vdw{} interactions are relevant only 
for neutral systems, we restrict the discussion to neutral 
hydrogen atoms (nuclear charge number $Z = 1$). 
In leading logarithmic approximations,
the Lamb shift Hamiltonian is approximated by~\cite{Je2003jpa}
\begin{equation} \label{eq:LambS}
H_{\mathrm{LS}}=\sum_{i=A,B}
\frac{4}{3}\,\alpha^2 m_e c^2\left(\frac{\hbar}{m_e c}\right)^3 \,
\ln\left(\alpha^{-2}\right) \,
\delta^{\left(3\right)}\left(\vec{r}_i\right).
\end{equation}

From Eq.~(6) of Ref.~\cite{AdEtAl2017vdWi},
we recall the \vdw{} Hamiltonian
\begin{align}
\label{HVDW}
H_{\rm vdW} =& \; \frac{e^2}{4\pi\epsilon_0} \,
\frac{  \vec r_A \cdot \vec r_B - 3 \,
(\vec r_A \cdot \hat R) \, 
(\vec r_B \cdot \hat R)}{R^3} \,,
\end{align}
where $\vec R = \vec R_A - \vec R_B$, $R = | \vec R|$ 
and $\hat{R}=\vec{R}/R$.
We shall assume that the hierarchy
\begin{equation}
\left< H_{\rm vdW} \right> \ll
\left< H_{\rm LS} \right> \ll
\left< H_{\rm FS} \right>
\end{equation}
is fulfilled for the entire distance range
relevant to the current investigation 
($R \gtrsim 30 \, a_0$).

We carefully distinguish different asymptotic 
regimes for the interatomic interaction.
In the so-called \vdw{} range of interatomic distances,
\begin{equation}
\label{vdWrange}
a_0 = \frac{\hbar}{\alpha m_e c} \ll 
R \ll 
\frac{\hbar}{\alpha^2 m_e c} = \frac{a_0}{\alpha} \,,
\end{equation}
the interatomic distance $R$ is much larger than the Bohr radius
$a_0 = \hbar/(\alpha m_e c)$, but much smaller than the wavelength 
$\approx a_0/\alpha$ 
of a typical optical transition, and the interaction 
is of the usual $R^{-6}$ functional form. 
This remains valid if one atom is in an excited $nS$ state.
In the so-called Casimir--Polder range,
\begin{equation}
\label{CPrange1}
R \gg \frac{\hbar}{\alpha^2 m_e c} = \frac{a_0}{\alpha} \,,
\end{equation}
the interatomic distance is much larger than the 
wavelength of an optical transition, and the interaction of 
\emph{ground-state} atoms has a $R^{-7}$ functional form.
For the long-range interaction involving excited metastable atoms, however,
we have to distinguish a third range of very large 
interatomic distances, 
\begin{equation}
\label{LSrange}
R \gg \frac{\hbar c}{\calL} \,,
\end{equation}
which we would like to refer to as the Lamb shift range
(where $\calL$ is a typical Lamb shift energy). 
Care is needed in the intermediate range
\begin{equation}
\frac{\hbar}{\alpha^2 \, m_e \, c} = \frac{a_0}{\alpha} \ll R \ll
\frac{\hbar c}{\calL} \,.
\end{equation}

A further complication arises.
The state with atom $A$ in the excited state
and atom $B$ in the ground state, 
$| nS \rangle_A \, | 1S \rangle_B$,
is degenerate with respect to 
the state $| 1S \rangle_A \, | nS \rangle_B$
with the quantum numbers reversed.
While there is no direct first-order coupling between 
the states due to the~\vdw{} interaction,
an off-diagonal term is obtained in second order.
It is of the same order-of-magnitude as the 
diagonal term, i.e., the term with the same 
in and out states. 
The Hamiltonian matrix in the 
basis of the degenerate states $| nS \rangle_A \, | 1S \rangle_B$ and
$| 1S \rangle_A \, | nS \rangle_B$ then has off-diagonal 
(``exchange'' or ``mixing'') terms
of second order in the \vdw{} interaction~\cite{Ch1972,AdEtAl2017vdWi}.

%
% Formalism
%
\subsection{Formalism for the Direct Terms}
\label{formalismA}

For $nS$--$1S$ interactions, the long-range interaction energy 
is the sum of three terms, namely, 
{\em (i)} a Wick-rotated interaction integral involving the 
nondegenerate states of the excited atom,
{\em (ii)} a Wick-rotated interaction with the 
quasi-degenerate states of the excited atom,
and {\em (iii)} the sum of pole terms,
due to lower-lying $mP$ states with $m \leq n-1$.
For details of the derivations, 
see Refs.~\cite{AdEtAl2017vdWi,JeDe2017vdw0,JeAdDe2017prl}.

First, we here restrict the discussion to the ``direct'' term
and indicate the specific contributions;
a summary of all contributing terms will be given in 
Sec.~\ref{add}.
The first contribution to the 
Wick-rotated term, involving nondegenerate states, 
is given as follows,
\begin{align}
\label{widetildeWdir}
{\widetilde \calW}^\dir(R) =& \;
-\frac{\hbar}{\pi \, c^4 \, (4 \pi \epsilon_0)^2} \,
\int\limits_0^\infty {\rm d}\omega\,
{\widetilde \alpha}_{nS}(\ii \omega) \,
\alpha_{1S}(\ii \omega)
\nonumber\\[0.1133ex]
& \; \times
{\rm e}^{-2\omega R/c}\,
\frac{\omega^4 \, }{R^2}
\left[ 1
+ 2\left(\frac{c}{\omega R}\right)
+ 5\left(\frac{c}{\omega R}\right)^2
\right.
\nonumber\\[0.1133ex]
& \; \left. + 6\left(\frac{c}{\omega R}\right)^3
+ 3\left(\frac{c}{\omega R}\right)^4
\right].
\end{align}
The nondegenerate contribution to the 
$nS$-state polarizability (denoted by a tilde) 
is given as
\begin{subequations}
\begin{align}
\label{alphanStilde}
\widetilde \alpha_{nS}(\omega) =& \;
\widetilde  P_{nS}(\omega) + \widetilde  P_{nS}(-\omega) \,,
\\[0.1133ex]
\label{PnStilde}
\widetilde P_{nS}(\omega) =& \; \frac{1}{3} \,
\sum_{m \neq n} 
\frac{\langle nS | \vec{d} | mP \rangle \cdot 
\langle mP | \vec{d} | nS \rangle }
{E_m - E_{nS} + \hbar \omega - \ii \, \epsilon} \,.
\end{align}
\end{subequations}
Here, $\vec d = e \, \vec r$ is the dipole operator.
The sum over $m$ includes the continuum 
states, and the sum over the magnetic quantum
numbers of the virtual $P$ states is implied.
However, note the restriction to nondegenerate states
in the sum over virtual states ($m \neq n$).
The ground-state polarizability is
\begin{subequations}
\begin{align}
\alpha_{1S}(\omega) =& \;
P_{1S}(\omega) + P_{1S}(-\omega) \,,
\\[0.1133ex]
\label{P1S}
P_{1S}(\omega) =& \; \frac{1}{3} \,
\sum_{m} 
\frac{\langle nS | \vec{d} | mP \rangle \cdot
\langle mP | \vec{d} | nS \rangle }
{E_m - E_{2S} + \hbar \omega - \ii \, \epsilon} \,.
\end{align}
\end{subequations}
The second Wick-rotated term, involving the degenerate states, 
is given as follows,
\begin{align}
\label{overlineWdir}
{\overline \calW}^\dir(R) =& \;
-\frac{\hbar}{\pi \, c^4 \, (4 \pi \epsilon_0)^2} \,
\int\limits_0^\infty {\rm d}\omega\,
{\overline \alpha}_{nS}(\ii \omega) \,
\alpha_{1S}(\ii \omega)
\nonumber\\[0.1133ex]
& \; \times
{\rm e}^{-2\omega R/c}\,
\frac{\omega^4 \, }{R^2}
\left[ 1
+ 2\left(\frac{c}{\omega R}\right)
+ 5\left(\frac{c}{\omega R}\right)^2
\right.
\nonumber\\[0.1133ex]
& \; \left. + 6\left(\frac{c}{\omega R}\right)^3
+ 3\left(\frac{c}{\omega R}\right)^4
\right].
\end{align}
Here, the degenerate part of the polarizability involves the 
$nP$ states, with the same principal quantum 
number as the reference state,
\begin{subequations}
\begin{align}
\label{alphanSbar}
\overline\alpha_{nS}(\omega) =& \;
\overline P_{nS}(\omega) + \overline P_{nS}(-\omega) \,,
\\[0.1133ex]
\label{PnSbar}
\overline P_{nS}(\omega) =& \; \frac{1}{9} \,
\frac{ \langle nS | \vec{d} | nP \rangle 
\cdot \langle nP | \vec{d} | nS \rangle }
{-{\calL}_n + \hbar \omega - \ii \, \epsilon}
\\[0.1133ex]
& \; + \frac{2 }{9} \,
\frac{ \langle nS | \vec{d} | nP \rangle 
\cdot \langle nP | \vec{d} | nS \rangle }
{{\calF}_n + \hbar \omega - \ii \, \epsilon },
\end{align}
\end{subequations}
where ${\calL}_n$  and ${\calF}_n$ are  the Lamb shift and
fine structure splittings between quasi-degenerate 
levels with principal quantum number $n$. 
(We have previously denoted by $\calL$ an energy 
commensurate with the Lamb shift energies 
$\calL_n$ in the range $2 \leq n \leq 12$.)
Explicitly,
\begin{subequations}
\begin{align}
{\calL}_n = E(nS_{1/2})-E(nP_{1/2})\, ,\\
{\calF}_n = E(nP_{3/2})-E(nS_{1/2})\,.
\end{align}
\end{subequations}
{Both the Lamb  shift, $\mathcal{L}_n$, and the fine structure
splitting,  $\mathcal{F}_n$,  decrease approximately as $1/n^3$ as the
principal quantum number $n$ increases~\cite{JeKoLBMoTa2005,hdel}.
The pole term~\cite{JeDe2017vdw0,JeAdDe2017prl} 
due to energetically lower $|m P\rangle$ states ($m < n$)
becomes
\begin{multline}
\calQ^\dir(R) =
- \frac{2}{3 (4 \pi \epsilon_0)^2 R^6}
\\[0.1133ex]
\times \sum_{m < n} \langle nS | \vec d | mP \rangle \cdot
\langle mP | \vec d | nS \rangle \,
\alpha_{1S}\left(\frac{E_{mn}}{\hbar}\right) 
\\[0.1133ex]
\times \exp\left(-\frac{2 \ii E_{mn} R}{\hbar c}\right)
\left[ 3 + 6 \ii \frac{E_{mn} R}{\hbar c} 
- 5 \left( \frac{E_{mn} R}{\hbar c} \right)^2 \right.
\\[0.1133ex]
\left. - 2 \ii \left( \frac{E_{mn} R}{\hbar c} \right)^3
+ \left( \frac{E_{mn} R}{\hbar c} \right)^4 \right] \,.
\end{multline}
The Schr\"{o}dinger energy difference is 
\begin{equation}
E_{mn} = -\frac{E_h}{2} \, 
\left( \frac{1}{m^2} - \frac{1}{n^2} \right) \,,
\end{equation}
where $E_h = \alpha^2 m_e c^2$ is the Hartree energy.
The real part of the pole term energy shift is 
\begin{multline}
\label{DirectPoleGeneral}
\calP^\dir(R) =
-\frac{2}{3 (4 \pi \epsilon_0)^2 R^6}
\\[0.1133ex]
\times \sum_{m < n} \langle nS | \vec d | mP \rangle \cdot
\langle mP | \vec d | nS \rangle \,
\alpha_{1S}\left(\frac{E_{mn}}{\hbar}\right)
\\[0.1133ex]
\times
\left\{\cos\left(\frac{2 E_{mn} R}{\hbar c}\right)  \,
\left( 3 - 5 \left(\frac{E_{mn} R}{\hbar c}\right)^2 + 
\left(\frac{E_{mn} R}{\hbar c}\right)^4 \right)
\right.
\\[0.1133ex]
\left.+ \frac{2 E_{mn} R}{\hbar c}
\sin\left( \frac{2 E_{mn} R}{\hbar c} \right) 
\left( 3 - \left(\frac{E_{mn} R}{\hbar c} \right)^2 \right) \right\}\,.
\end{multline}
The corresponding width term $\Gamma^\dir(R)$
is obtained from the relation
\begin{equation}
\calQ^\dir(R) = \calP^\dir(R) - \frac{\ii}{2} \Gamma^\dir(R)  
\end{equation}
and reads
\begin{multline}
\Gamma^\dir(R) =
-\frac{4}{3 (4 \pi \epsilon_0)^2 R^6}
\\[0.1133ex]
\times \sum_{m < n} \langle nS | \vec d | mP \rangle \cdot
\langle mP | \vec d | nS \rangle \,
\alpha_{1S}\left(\frac{E_{mn}}{\hbar}\right)
\\[0.1133ex]
\times
\left\{ 
\sin\left(\frac{2 E_{mn} R}{\hbar c}\right)  \,
\left( 3 - 5 \left(\frac{E_{mn} R}{\hbar c}\right)^2 +
\left(\frac{E_{mn} R}{\hbar c}\right)^4 \right)
\right.
\\[0.1133ex]
\left. - \frac{2 E_{mn} R}{\hbar c}
\cos\left( \frac{2 E_{mn} R}{\hbar c} \right)
\left( 3 - \left(\frac{E_{mn} R}{\hbar c} \right)^2 \right) \right\}\,.
\end{multline}
%
% We recognize a number of prefactors 
% which were also present in Eq.~(19) of Ref.~\cite{SaKa2015}.

%
% Formalism for the Mixing Terms
%
\subsection{Formalism for the Mixing Terms}
\label{formalismB}

Just as for the direct term, we need to identify 
a nondegenerate contribution ${\widetilde \calW}^\mix(R)$
to the Wick-rotated term,
a degenerate contribution ${\overline \calW}^\mix(R)$,
and pole term $\calP^\mix(R)$.
The first Wick-rotated term, involving nondegenerate states, 
is given as follows,
\begin{align}
{\widetilde \calW}^\mix(R) =& \;
-\frac{\hbar}{\pi \, c^4 \, (4 \pi \epsilon_0)^2} \,
\int\limits_0^\infty {\rm d}\omega\,
{\widetilde \alpha}_{\underline{nS}1S}(\ii \omega) \,
\alpha_{nS\underline{1S}}(\ii \omega)
\nonumber\\[0.1133ex]
& \; \times
{\rm e}^{-2\omega R/c}\,
\frac{\omega^4 \, }{R^2}
\left[ 1
+ 2\left(\frac{c}{\omega R}\right)
+ 5\left(\frac{c}{\omega R}\right)^2
\right.
\nonumber\\[0.1133ex]
& \; \left. + 6\left(\frac{c}{\omega R}\right)^3
+ 3\left(\frac{c}{\omega R}\right)^4
\right].
\end{align}
The mixed polarizabilities are given as
\begin{subequations}
\begin{align}
\label{pola}
\widetilde \alpha_{\underline{nS}1S}(\omega) =& \;
\frac13 \sum_{m \neq n} \sum_\pm
\frac{\langle nS | \vec{d} | mP \rangle \cdot 
\langle mP | \vec{d} | 1S \rangle }
{E_m - E_{nS} \pm \hbar \omega - \ii \, \epsilon} \,,
\\[0.1133ex]
\label{polb}
\alpha_{nS\underline{1S}}(\omega) =& \;
\frac13 \sum_{m} \sum_\pm
\frac{\langle nS | \vec{d} | mP \rangle \cdot
\langle mP | \vec{d} | 1S \rangle }
{E_m - E_{1S} \pm \hbar \omega - \ii \, \epsilon} \,.
\end{align}
\end{subequations}
Note the restriction to nondegenerate states ($m \neq n$)
in the sum over virtual states, in the expression for 
$\widetilde \alpha_{\underline{nS}1S}(\omega)$.
The second Wick-rotated term, involving the degenerate states, 
is given as follows,
\begin{align}
{\overline \calW}^\mix(R) =& \;
-\frac{\hbar}{\pi \, c^4 \, (4 \pi \epsilon_0)^2} \,
\int\limits_0^\infty {\rm d}\omega\,
{\overline \alpha}_{\underline{nS}1S}(\ii \omega) \,
\alpha_{nS\underline{1S}}(\ii \omega)
\nonumber\\[0.1133ex]
& \; \times
{\rm e}^{-2\omega R/c}\,
\frac{\omega^4 \, }{R^2}
\left[ 1
+ 2\left(\frac{c}{\omega R}\right)
+ 5\left(\frac{c}{\omega R}\right)^2
\right.
\nonumber\\[0.1133ex]
& \; \left. + 6\left(\frac{c}{\omega R}\right)^3
+ 3\left(\frac{c}{\omega R}\right)^4
\right].
\end{align}
Here, the degenerate part of the polarizability involves the 
$nP$ states, with the same principal quantum 
number as the reference state,
\begin{align}
\label{alphanS1Sbar}
\overline\alpha_{\underline{nS}1S}(\omega) =& \;
\frac{1}{9} \,
\frac{ \langle nS | \vec{d} | nP \rangle
\cdot \langle nP | \vec{d} | 1S \rangle }
{-{{\calL}_n} + \hbar \omega - \ii \, \epsilon}
\nonumber\\[0.1133ex]
& \; + \frac{2 }{9} \,
\frac{ \langle nS | \vec{d} | nP \rangle
\cdot \langle nP | \vec{d} | 1S \rangle }
{{\calF}_n + \hbar \omega - \ii \, \epsilon } \,.
\end{align}

The mixed pole term due to 
energetically lower $|m P\rangle$ states becomes
\begin{multline}
\calQ^\mix(R) =
-\frac{2}{3 (4 \pi \epsilon_0)^2 R^6}
\\[0.1133ex]
\times \sum_{m < n} \langle nS | \vec d | mP \rangle \cdot
\langle mP | \vec d | 1S \rangle \,
\alpha_{nS\underline{1S}}\left(\frac{E_{mn}}{\hbar}\right) 
\\[0.1133ex]
\times \exp\left(-\frac{2 \ii E_{mn} R}{\hbar c}\right)
\left[ 3 + 6 \ii \frac{E_{mn} R}{\hbar c} 
- 5 \left( \frac{E_{mn} R}{\hbar c} \right)^2 \right.
\\[0.1133ex]
\left. - 2 \ii \left( \frac{E_{mn} R}{\hbar c} \right)^3
+ \left( \frac{E_{mn} R}{\hbar c} \right)^4 \right] \,.
\end{multline}
The real part is 
\begin{multline}
\label{Pmix}
\calP^\mix(R) =
-\frac{2}{3 (4 \pi \epsilon_0)^2 R^6}
\\[0.1133ex]
\times \sum_{m < n} \langle nS | \vec d | mP \rangle \cdot
\langle mP | \vec d | 1S \rangle \,
\alpha_{nS\underline{1S}}\left(\frac{E_{mn}}{\hbar}\right)
\\[0.1133ex]
\times
\left\{\cos\left(\frac{2 E_{mn} R}{\hbar c}\right)  \,
\left( 3 - 5 \left(\frac{E_{mn} R}{\hbar c}\right)^2 + 
\left(\frac{E_{mn} R}{\hbar c}\right)^4 \right)
\right.
\\[0.1133ex]
\left.+ \frac{2 E_{mn} R}{\hbar c}
\sin\left( \frac{2 E_{mn} R}{\hbar c} \right) 
\left( 3 - \left(\frac{E_{mn} R}{\hbar c} \right)^2 \right) \right\}\,.
\end{multline}
The corresponding width term $\Gamma^\mix(R)$
is obtained from the relation
\begin{equation}
\calQ^\mix(R) = \calP^\mix(R) - \frac{\ii}{2} \Gamma^\mix(R)
\end{equation}
and reads 
\begin{multline}
\Gamma^\mix(R) =
-\frac{4}{3 (4 \pi \epsilon_0)^2 R^6}
\\[0.1133ex]
\times \sum_{m < n} \langle nS | \vec d | mP \rangle \cdot
\langle mP | \vec d | 1S \rangle \,
\alpha_{nS\underline{1S}}\left(\frac{E_{mn}}{\hbar}\right)
\\[0.1133ex]
\times
\left\{ 
\sin\left(\frac{2 E_{mn} R}{\hbar c}\right)  \,
\left( 3 - 5 \left(\frac{E_{mn} R}{\hbar c}\right)^2 +
\left(\frac{E_{mn} R}{\hbar c}\right)^4 \right)
\right.
\\[0.1133ex]
\left. - \frac{2 E_{mn} R}{\hbar c}
\cos\left( \frac{2 E_{mn} R}{\hbar c} \right)
\left( 3 - \left(\frac{E_{mn} R}{\hbar c} \right)^2 \right) \right\}\,.
\end{multline}

%
% Adding direct and mixed terms
%
\subsection{Adding direct and mixed terms}
\label{add}

Depending on the symmetry of the two-atom wave function,
we have for the eigenenergies of the two-atom 
system~\cite{Ch1972,AdEtAl2017vdWi}
\begin{subequations}
\begin{align}
E(R) =& \; E^\dir(R) \pm E^\mix(R) \,,
\\[0.1133ex]
E^\dir(R) =& \; 
{\widetilde {\calW}}^\dir(R) + {\overline {\calW}}^\dir(R) + \calQ^\dir(R) \,,
\\[0.1133ex]
E^\mix(R) =& \; 
{\widetilde {\calW}}^\mix(R) + {\overline {\calW}}^\mix(R) + 
\calQ^\mix(R) \,.
\end{align}
\end{subequations}
For the real part of the interaction energy, 
\begin{equation}
{\rm Re} \, E(R) =
{\rm Re} \, E^\dir(R) \pm {\rm Re} \, E^\mix(R) \,,
\end{equation}
one has
\begin{subequations}
\begin{align}
{\rm Re} \, E^\dir(R) =& \; 
\calW^\dir(R) + \calP^\dir(R) \nonumber\\
=& \; 
{\widetilde {\calW}}^\dir(R) + {\overline {\calW}}^\dir(R) +
\calP^\dir(R) \,,
\\[0.1133ex]
{\rm Re} \, E^\mix(R) =& \; 
 \calW^\mix(R) + \calP^\mix(R) \nonumber\\
=& \; 
{\widetilde {\calW}}^\mix(R) + {\overline {\calW}}^\mix(R) +
\calP^\mix(R) \,.
\end{align}
\end{subequations}
The sign of the mixing term depends on the 
symmetry of the wave function of the two-atom system~\cite{Ch1972}.
In the following, we will concentrate on
the real part of the energy shift and 
use the symbols $E(R)$ and ${\rm Re} \, E(R)$ synonymously,
for both the direct as well as the mixing terms.

%
% $\bf{1S}$--$\bf{2S}$ Long--Range Interaction
%
\section{\texorpdfstring{$\maybebm{3S}$}{3S}--\texorpdfstring{$\maybebm{1S}$}{1S}
Interaction}
\label{sec3}

%
% van der Waals range of interatomic distance}
%
\subsection{\VDW{} range} 
\label{subsec:CloseD}

In the \vdw{} distance range~\eqref{vdWrange},
\begin{equation}\label{vdwRegime}
a_0 \ll R \ll \frac{a_0}{\alpha} \,,
\end{equation}
the interaction is nonretarded, and the 
interaction energy is well approximated by the form
\begin{equation}
\label{C6D6M6}
E(R) \approx -\frac{C_6}{R^6} = -\frac{D_6 \pm M_6}{R^6} \,.
\end{equation}
The \vdw{} coefficient $C_6 = D_6 \pm M_6$
contains a direct term $D_6$ and a mixing coefficient $M_6$.

First, we focus on the direct term.
According to Sec.~\ref{formalismA},
$D_6$ is the sum of a nondegenerate Wick-rotated term
$\widetilde D_6$,
a degenerate Wick-rotated term $\overline D_6$,
and a pole contribution $D_6^\calP$. 
One writes
\begin{equation}
D_6(3S; 1S) = \widetilde{D}_6(3S; 1S) +
\overline D_6(3S; 1S) + D_6^\calP(3S; 1S) \,.
\end{equation}
Let us start with the nondegenerate contribution
\begin{equation}
\label{C6int}
\widetilde {D}_6(3S; 1S) =
\frac{3}{\pi} \frac{\hbar}{(4 \pi \epsilon_0)^2} 
\int_0^\infty {\rm d}\omega\, 
\widetilde \alpha_{3S}(\ii \omega) \, 
\alpha_{1S}(\ii \omega) \,,
\end{equation}
where $\widetilde \alpha_{3S}(\omega)$ has been 
defined in Eq.~\eqref{alphanStilde}.
For the $1S$ polarizability, the result was recently 
given in Eqs.~(15),~(27a) and~(27b) of Ref.~\cite{AdEtAl2017vdWi}.
For the $3S$ state, one obtains the nondegenerate matrix element
\begin{align}
\label{tildeP3S}
\widetilde P_{3S}(\omega) =&\frac{e^2a_0^2}{E_h}
\left[ \frac{54 \tau^2}{(1- \tau )^8(1+ \tau )^6} \right.\left(15538 \tau ^{12}- 
2852 \tau ^{11}\right. \nonumber\\
-&13283  \tau ^{10} +2090 \tau ^9+ 2871 \tau ^8 +
40  \tau ^7 -62 \tau ^6 \nonumber\\
-&\left. 492  \tau ^5 +128 \tau ^4 + 236 \tau ^3 -
95  \tau ^2 -46 \tau  +23\right)  \nonumber  \\
+ &\frac{6912 \, \tau ^{9}}{(1- \tau )^8(1+ \tau )^8} \, 
(-1+9 \tau ^2)(3-7 \tau ^2)^2 \nonumber  \\
\times& \left.  {}_2F_{1}\left(1, -3 \tau, 1-3 \tau, 
\frac{(1-  \tau )^2}{(1+ \tau )^2}\right) -
\frac{972\ \tau ^2}{1-\tau ^2} \right], \nonumber\\
& \tau = \left( 1+\frac{18\hbar\omega}{\alpha^2 m_e c^2}\right)^{-1/2}.    
\end{align}
%%%%%%
%
The virtual $3P$ state is excluded from the sum over states in 
Eq.~\eqref{tildeP3S} by the explicit subtraction of the term 
$972 \tau^2/(1-\tau^2) = 54 E_h/(\hbar \omega)$.
One can verify that the expression~\eqref{tildeP3S}
is finite in the limit $\tau \to 1$, which is equivalent to 
vanishing photon energy $\omega \to 0$.

The polarizability $\widetilde \alpha_{3S}(\omega)$ is recovered
according to Eq.~\eqref{alphanStilde}, namely,
\begin{equation}
\label{polarizability3S}
\widetilde \alpha_{3S}(\omega) = 
\widetilde P_{3S}(\omega) +\widetilde P_{3S}(-\omega) \,.
\end{equation}
A numerical evaluation of Eq.~\eqref{C6int} leads to 
the result
\begin{equation}
\label{res1}
\widetilde D_6(3S; 1S) = 180.320\,073\,947\,E_h\,a_0^6 \,.
\end{equation}

\begin{figure}[t!]
\begin{center}
\begin{minipage}{0.91\linewidth}
\begin{center}
\includegraphics[width=0.91\linewidth]{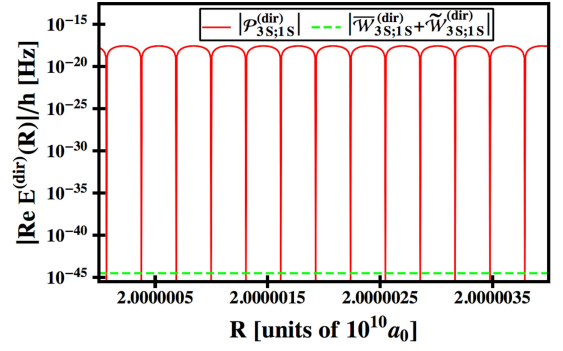}
\end{center}
\end{minipage}
\caption{\label{fig1} Interaction energy in the $3S$--$1S$ system
as a function of the interatomic distance, $R$,
for very-long range. The pole term dominates
over the Casimir-Polder term.
However, the overall magnitude of the interaction is very small.}
\end{center}
\end{figure}

\begin{figure}[t!]
\begin{center}
\begin{minipage}{0.91\linewidth}
\begin{center}
\includegraphics[width=0.91\linewidth]{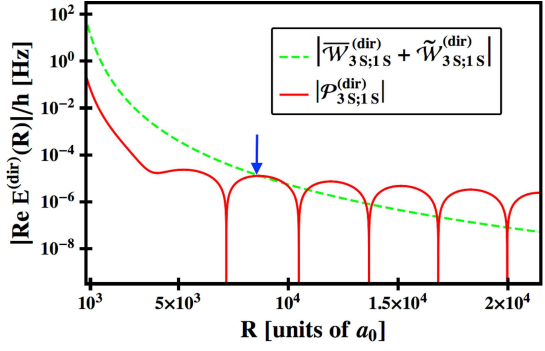}
\end{center}
\end{minipage}
\caption{\label{fig2} Interaction energy in the $3S$--$1S$ system 
as a function of the interatomic distance $R$
in the intermediate range. Initially, the Wick-rotated term 
dominates the pole term. As the interatomic distance increases,
the pole term gradually dominates. 
The arrow indicates the position where the pole term becomes 
comparable to  the Wick-rotated term.
Note the logarithmic scale of the ordinate axis;
the logarithm tends to $-\infty$ upon a sign 
change of the interaction energy.}
\end{center}
\end{figure}

The degenerate contribution to $D_6$ can be handled analytically. It reads
\begin{equation}
\label{defoverlineC6}
{\overline D}_6(3S;1S) =
\frac{3}{\pi} \, \frac{\hbar}{(4 \pi \epsilon_0)^2}
\int\limits_0^\infty {\rm d}\omega \, 
\overline\alpha_{3S}(\ii \omega) \, \alpha_{1S}(0) \,.
\end{equation}
Using Eqs.~(23) and~(24) of Ref.~\cite{AdEtAl2017vdWi},
one easily obtains the result for ${\overline D}_6(3S;1S)$,
\begin{equation} 
\label{res2}
{\overline D}_6(3S;1S) = 729 \, E_h\, a_0^6 \,,
\end{equation} 
In the calculation, one uses the well-known result 
\begin{equation}
\label{static1S}
\alpha_{1S}\left(0\right) = \frac92 \frac{e^2 a_0^2}{E_h} 
\end{equation}
for the static polarizability of hydrogen.

In the \vdw{} range, the pole term given in
Eq.~\eqref{DirectPoleGeneral} can be approximated as 
\begin{multline}
\label{DirectPole:closerange}
\calP^\dir(R) \approx
-\frac{2}{ (4 \pi \epsilon_0)^2 R^6}
\\[0.1133ex]
\times \sum_{m < n} \langle nS | \vec d | mP \rangle \cdot
\langle mP | \vec d | nS \rangle \,
\alpha_{1S}\left(\frac{E_{mn}}{\hbar}\right)\,,
\end{multline}
in view of the fact that $E_{mn} \, R/(\hbar c) 
\sim \alpha R/a_0 \to 0$.
For the $3S$--$1S$ interaction, this implies that 
the pole term yields another nontrivial contribution
to $D_6$, which can be expressed as follows,
\begin{align}\label{DP3S1S}
D_6^\calP(3S;1S) =&\frac{2}{(4\pi\epsilon_0)^2 }
 \langle 3S|\vec{d}| 2P\rangle\cdot \langle 2P|\vec{d}|3S\rangle \nonumber\\
\times& \alpha_{1S}\left( \omega= \frac{E_{2P3S}}{\hbar} \right) \,,
 \end{align}
where the sum over magnetic projections of the virtual 
$2P$ state is implied,
\begin{equation}
\left|\langle 3S | \vec{d} | 2P\rangle\right|^2 =  
\frac{2^{15}\times 3^8 }{5^{12}} \, e^2 a_0^2 \,.
\end{equation}
The polarizability $\alpha_{1S}\left( \omega= E_{2P3S}/\hbar \right)$ slightly
differs from the static value given in Eq.~\eqref{static1S},
\begin{equation}
\label{alpha1S:2sS3P}
\alpha_{1S}\left( \omega= \frac{E_{2P3S}}{\hbar} \right)= 
4.632\,338\,310\, \frac{e^2 a_0^2}{E_h}.
\end{equation}
Thus, the direct pole term, $D_6^\calP(3S;1S)$,  is given by 
\begin{equation}
\label{res3}
D_6^\calP(3S;1S) = 8.158\,497\,516\,E_h\,a_0^6 \,.
\end{equation}
Adding the results from Eq.~\eqref{res1},~\eqref{res2} and~\eqref{res3},
one finally obtains the complete result for the $D_6$ coefficient
of the $3S$--$1S$ interaction,
\begin{align}
\label{C62S1Sshortnum}
D_6(3S; 1S) =& \; 
\widetilde{D}_6(3S; 1S) +\overline D_6(3S; 1S) +
D^\calP_6(3S; 1S) 
\nonumber\\[0.1133ex]
%=& \; 917.478\,571\,463\,894\,532 \, E_h\, a_0^6 \,,
=& \; 917.478\,571\,464 \, E_h\, a_0^6 \,.
\end{align}

We have  verified the result~\eqref{C62S1Sshortnum} 
by two alternative numerical methods.
A discrete lattice representation of the 
radial Schr\"{o}dinger equation and its spectrum
(Ref.~\cite{SaOe1989}) can be used in order to 
approximate the radial component of the 
Schr\"{o}dinger--Coulomb propagator.
This leads to an alternative evaluation of the 
$D_6$ coefficient in terms of an explicit sum over virtual 
states comprising the pseudo-spectrum (see Ref.~\cite{DeYo1973}).
The result confirms that $D_6(3S;1S) = 917.478(1)$.
Another possibility to verify the result~\eqref{C62S1Sshortnum}
consists in an approach based on 
``intermediate quantum numbers'', as
outlined in the text surrounding Eq.~(33) of Ref.~\cite{AdEtAl2017vdWi}.
The basic idea is that one can shift the reference-state
quantum numbers artificially in the integrals 
describing the \vdw{} energy, 
provided the bound-state energies of both involved
states combined add up to the same total
reference-state energy in the two-atom system.
This second approach also confirms the result~\eqref{C62S1Sshortnum}.

%
% table 1
%
\begin{table*}[!thbp]
\begin{center}
\begin{minipage}{0.9\linewidth}
\begin{center}
\caption{\label{table1}
Numerical values of the degenerate contributions $\overline{D}_6$,
nondegenerate contributions $\widetilde D_6$,  and the  pole term contributions
$D^\calP_6$  to the direct $D_6$  van der Waals coefficients for two-atom
systems in the van der Waals range. The coefficients are given in units of
$E_h\,a_0^6$.} 
\begin{tabular}{c@{\hspace*{0.3cm}}cS[table-format=6.9e0]S[table-format=6.9e0]S[table-format=6.9e0]}
\hline
\hline
\rule[-2mm]{0mm}{5.5mm}
System & 
\multicolumn{1}{c}{$\overline D_6$} &
\multicolumn{1}{c}{$\widetilde D_6$} & 
\multicolumn{1}{c}{$D^\calP_6$} &
\multicolumn{1}{c}{${D}_6=\overline D_6+\widetilde D_6+D^\calP_6$} \\
\hline
\rule[-2mm]{0mm}{5.5mm}
$3S$--$1S$\ &729      &  180.320073947     &    8.158497516  &  917.478571464\\
\rule[-2mm]{0mm}{5.5mm}
$4S$--$1S$\ &\  2430  &  415.867781719     &  55.313793349   &  2901.174002323\\
\rule[-2mm]{0mm}{5.5mm}
$5S$--$1S$\ &\  6075  &  797.620619336     &  199.631309750  &  7072.251929086\\
\rule[-2mm]{0mm}{5.5mm}
$6S$--$1S$ &  $\tfrac{25515}{2}$         &  1361.858822274 &  526.146484053  &  14615.505306328\\
\rule[-2mm]{0mm}{5.5mm}
$7S$--$1S$ &  23814 &   2144.976599069   &  1146.872740254 &  27105.849339323\\
\rule[-2mm]{0mm}{5.5mm}
$8S$--$1S$ &  40824 &  3183.421765600    &  2200.822886660 &  46208.244652261\\
\rule[-2mm]{0mm}{5.5mm}
$9S$--$1S$ &  65610 &  4513.658548391    &  3854.012517378 &  73977.671065769\\
\rule[-2mm]{0mm}{5.5mm}
$10S$--$1S$ &   $\tfrac{200475}{2}$      &  6172.157501976 &  6299.459903847 &  112709.117405823\\
\rule[-2mm]{0mm}{5.5mm}
$11S$--$1S$ &  147015 &  8195.391734362  &  9757.185355911 &  164967.577090273\\
\rule[-2mm]{0mm}{5.5mm}
$12S$--$1S$ &  208494 &  10619.835391823 &  22062.734967733 &  241176.570359557\\
\hline 
\hline
\end{tabular}
\end{center}
\end{minipage}
\end{center}
\end{table*}

Similarly, the mixing term $M_6$ is obtained as the sum 
of a Wick-rotated nondegenerate term $ \widetilde{M}_6(3S; 1S)$,
a  Wick-rotated degenerate contribution $\overline{M}_6(3S; 1S)$,
and a pole term $M^\calP_6(3S; 1S)$,
\begin{equation}
M_6(3S; 1S) = \widetilde{M}_6(3S; 1S) +\overline{M}_6(3S; 1S) +
M^\calP_6(3S; 1S) \,.
\end{equation}
The nondegenerate Wick-rotated contribution 
\begin{equation}
\widetilde{M}_6(3S; 1S)=\frac{3}{\pi} \frac{\hbar}{(4 \pi \epsilon_0)^2} 
\int\limits_0^\infty {\rm d}\omega\,
{\widetilde \alpha}_{\underline{3S}1S}(\ii \omega) \,
\alpha_{3S\underline{1S}}(\ii \omega) 
\end{equation}
is evaluated numerically, which yields 
\begin{equation}
\widetilde{M}_6(3S; 1S)= -5.588\,159\,518\,  E_h\, a_0^6\, .
\end{equation}
The degenerate coefficient $\overline{M}_6(3S; 1S)$ is given by 
\begin{align}
\label{OverlineM3s1s}
\overline{M}_6(3S; 1S)=& \;
\frac{3}{\pi}\frac{\hbar}{(4 \pi \epsilon_0)^2} \,
\int\limits_0^\infty {\rm d}\omega\,
{\overline \alpha}_{\underline{3S}1S}(\ii \omega) \,
\alpha_{3S\underline{1S}}(0) \,,
\end{align}
where we refer to Eq.~\eqref{alphanS1Sbar}
for the definition of ${\overline \alpha}_{\underline{3S}1S}(\omega)$.
We first carry out the integration and then take the limit 
${\calL}_3\to 0$, ${\calF}_3\to 0$ 
at the end of the calculation. 
The product of the matrix element of the dipole moment operators
(with the sum over the magnetic projections implied) is
\begin{align}
\langle 1S| \vec{d} | 3P \rangle \cdot \langle 3P| \vec{d} | 3S\rangle=
-\frac{243 \sqrt{3}\, e^2 a_0^2}{64} \,.
\end{align}
The mixed static polarizability
$\alpha_{3S\underline{1S}}(0)$ is given by 
\begin{align}
\alpha_{3S\underline{1S}}(0) = -\frac{621\sqrt{3}\, e^2 a_0^2}{512 \, E_h}.
\end{align}
An analytic calculation of the integral in Eq.~\eqref{OverlineM3s1s}
thus gives the result
\begin{align}
\label{OverlineM3s1sFinal}
\overline{M}_6(3S; 1S)=& \;
\frac{3^9\times 23 }{2^{15}}\, E_h\, a_0^6 
= 13.815\, 582\,275\, E_h\, a_0^6 \, .
\end{align}
Similar to the direct 
pole term, the mixing pole term  $M^\calP_6(3S; 1S)$ is given by 
\begin{multline}\label{MP3S1S}
M_6^\calP(3S;1S) = \frac{2}{(4\pi\epsilon_0)^2 }
\langle 1S|\vec{d}| 2P\rangle\cdot \langle 2P|\vec{d}|3S\rangle 
\\[0.1133ex]
\times \alpha_{3S\underline{1S}}\left( \omega= \frac{E_{2P3S}}{\hbar} \right) 
\\[0.1133ex]
= \frac{2}{(4\pi\epsilon_0)^2 } \, \left(-2.159\,394\,993\right) \, \frac{e^2a_0^2}{E_h} 
\, \frac{2^{15}e^2a_0^2}{5^6\times\sqrt{3}}
\\[0.1133ex]
=-5.229\,153\,219\, E_h\, a_0^6 \, .
\end{multline}
Thus, the total mixing contribution $M_6(3S;1S)$ is 
\begin{align}
\label{M62S1Sshortnum}
M_6(3S;1S) =& \; 
\widetilde{M}_6(3S; 1S)+ \overline{M}_6(3S; 1S) + M_6^\calP(3S;1S)
\nonumber\\
=& \; 2.998\,269\,538\, E_h\, a_0^6 \, .
\end{align}
While the direct and mixing coefficients are of the 
same order-of-magnitude for $2S$--$1S$ interactions
($\approx 176.75$ versus $\approx 27.98$), they
differ by two orders-of-magnitude in the case of the 
$3S$--$1S$ system.

%
% table2
% 
\begin{table*}[!thbp]
\begin{center}
\begin{minipage}{0.9\linewidth}
\begin{center}
\caption{\label{table2}
Numerical values of the degenerate contributions $\overline{M}_6$,
nondegenerate contributions $\widetilde M_6$,  and the  pole term contributions
$M^\calP_6$  to the mixed $M_6$  van der Waals coefficients for two-atom
systems in the van der Waals range. The coefficients are given in units of
$E_h\,a_0^6$.} 
\begin{tabular}{c@{\hspace*{0.3cm}}cS[table-format=-6.9e0]S[table-format=-6.9e0]S[table-format=-6.9e0]}
\hline
\hline
\multicolumn{1}{c}{\rule[-2mm]{0mm}{5.5mm}System} &
\multicolumn{1}{c}{$\overline M_6$} &
\multicolumn{1}{c}{$\widetilde M_6$} &
\multicolumn{1}{c}{$M^\calP_6$} &
\multicolumn{1}{c}{${M}_6=\overline M_6+\widetilde M_6+M^\calP_6$}
\\
\hline
\rule[-2mm]{0mm}{5.5mm}
$3S$--$1S$  &   13.815582275  &  -5.588159518 &  -5.229153219 &  2.998269538\\
\rule[-2mm]{0mm}{5.5mm}
$4S$--$1S$  &   8.015439766 &  -3.063629332 &  -4.033187464 &  0.918622970\\
\rule[-2mm]{0mm}{5.5mm}
$5S$--$1S$  &   5.716898855 &  -2.006704605 &  -3.302240659 &  0.407953591\\
\rule[-2mm]{0mm}{5.5mm}
$6S$--$1S$  &   4.480588908 &  -1.435991892 &  -2.825817540  &  0.218779478\\
\rule[-2mm]{0mm}{5.5mm}
$7S$--$1S$  &   3.702266657 &   -1.085159560 &  -2.485383226 &  0.131723872\\
\rule[-2mm]{0mm}{5.5mm}
$8S$--$1S$  &   3.163734811 &  -0.851710237 &  -2.224628639 &  0.087395934\\
\rule[-2mm]{0mm}{5.5mm}
$9S$--$1S$  &   2.767122768 &  -0.687554678 &  -2.020512545 &  0.059055545\\
\rule[-2mm]{0mm}{5.5mm}
$10S$--$1S $ &  2.461858057 &   -0.567328345 &  -1.851944579 &  0.042585133\\
\rule[-2mm]{0mm}{5.5mm}
$11S$--$1S$  &   2.219074417 &  -0.476448189 &  -1.711091985 &  0.031534242\\
\rule[-2mm]{0mm}{5.5mm}
$12S$--$1S$  &   2.021036738 &  -0.405984611 &  -1.590955009 &  0.024097118\\
\hline 
\hline
\end{tabular}
\end{center}
\end{minipage}
\end{center}
\end{table*}

%
% Intermediate Distances
%
\subsection{Intermediate distance} 
\label{subsec:MidD}

In the intermediate range of interatomic distances,
\begin{equation}
\label{intermediate}
\frac{a_0}{\alpha} \ll R \ll \frac{\hbar c}{{\calL}} \,,
\end{equation}
the treatment becomes a little sophisticated.
As far as $\widetilde {\calW}$ is concerned, we are 
in the Casimir--Polder regime where the result 
is given by an $R^{-7}$ interaction.
However, we incur a contribution proportional 
to $R^{-6}$ from the quasi-degenerate $3P$ 
states, i.e., from $\overline {\calW}$.
This contribution competes with the oscillatory
long-range tails from the pole terms,
which eventually drop off only as $1/R^2$.

From the quasi-degenerate states,
using the approach outlined in Eqs.~(23) and~(24) 
of Ref.~\cite{AdEtAl2017vdWi}, one obtains
\begin{align}
\label{C62S1Smedium}
\overline {\calW}_{3S; 1S}^{\left(\mathrm{dir}\right)}\left(R\right) 
=& \; - \frac{\overline D_6(3S; 1S)}{R^6} 
= - 729 \,E_h \left( \frac{a_0}{R} \right)^6 \,.
\end{align}
The Wick-rotated contribution to 
the interaction is thus still of the $R^{-6}$ form, as it is in the \vdw{} 
range, but the coefficient is reduced in magnitude as compared to 
Eq.~\eqref{C62S1Sshortnum}.
In the intermediate range,
the nondegenerate contribution
$\widetilde {\calW}_{3S; 1S}^{\left(\mathrm{dir}\right)}\left(R\right) $
is much smaller than
$\overline {\calW}_{3S; 1S}^{\left(\mathrm{dir}\right)}\left(R\right)$;
it follows a $1/R^7$ law.

Let us now look into the pole term contribution, 
$\mathcal{P}_{3S;1S}^{(\mathrm{dir})}(R)$, in the Casimir-Polder range.
The Wick rotation from the positive real axis onto the imaginary
axis picks up two  poles at $\omega= -E_{3P_{1/2},3S}/\hbar +  \ii \epsilon = 
-\calL_3 +  \ii \epsilon$  
and $\omega= -E_{2P,3S}/\hbar + i\epsilon$, which, respectively, are due to
the presence of the quasi degenerate $3P_{1/2}$ level and the low lying $2P$ level.  
The contribution of the quasi degenerate $3P_{1/2}$ level to the 
$1/R^6$-part of the pole term is 
already contained in Eq.~\eqref{C62S1Smedium}; 
we observe that the $1/R^6$ term in Eq.~\eqref{DirectPoleGeneral} does not 
have additional factors of $E_{mn} \sim \calL_3$
and is therefore not suppressed by the Lamb shift numerators.
By contrast,
the terms of order $1/R^4$ and $1/R^2$ in Eq.~\eqref{DirectPoleGeneral}
from the pole term due to the $3P_{1/2}$ levels
are suppressed by the very small energy factor 
in the numerators of Eq.~\eqref{DirectPoleGeneral}
(proportional to $\calL_3^2$ and $\calL_3^4$,
respectively). For simplicity, we treat the contribution 
of the $3P_{1/2}$ and $3P_{3/2}$ levels uniformly 
by assigning their contribution to the 
Wick-rotated term. This procedure follows the one 
adopted in Eqs.~(45a), (45b) and~(46) of Ref.~\cite{AdEtAl2017vdWi}.

Finally, the direct pole term for the $3S$--$1S$ system 
(contribution of the lower-lying $2P$ states) reads 
\begin{multline}\label{Pole3S1SCP}
\mathcal{P}_{3S;1S}^{(\mathrm{dir})}(R) =
- \frac{2^{15}\times 3^8}{5^{12}} \, \frac{E_h a_0^6}{R^6}\,  
\alpha^{\rm dl}_{1S}\left(\frac{5\,E_h}{72\hbar}\right) 
\\
\times\left\{ \cos\left( \frac{5 \alpha R}{36\, a_0} \right)
\left[ 3-5 \left(\frac{5 \alpha R}{72\, a_0} \right)^2 
+ \left(\frac{5 \alpha R}{72\, a_0}\right)^4 \right] \right.
\\
+ \left. \frac{5 \alpha R}{36\, a_0} \, 
\sin \left(\frac{5 \alpha R}{36\, a_0}\right)
\left[ 3 - \left(\frac{5 \alpha R}{72\, a_0}\right)^2 \right]
\right\} \,.
\end{multline}
Here, $\alpha^{\rm dl}_{1S}( \omega )$
is the dimensionless polarizability, which we define as follows,
\begin{equation}
\alpha^{\rm dl}_{1S}( \omega ) =
\frac{E_h}{e^2 \, a_0^2} \;
\alpha_{1S}( \omega )  \,.
\end{equation}
For clarity, we add that our dimensionless polarizability
could otherwise be characterized as the 
numerical value of the 
atomic polarizability in atomic units.

The Wick-rotated part of the mixing term is the sum 
\begin{equation}
{\calW}_{3S;1S}^\mix(R) = 
{\widetilde {\calW}}_{3S;1S}^\mix(R) +
{\overline {\calW}}_{3S;1S}^\mix(R) \,.
\end{equation}
The nondegenerate part 
${\widetilde {\calW}}_{3S;1S}^\mix(R)$
follows an $R^{-7}$ power law, 
whereas the degenerate part 
${\overline {\calW}}_{3S;1S}^\mix(R)$
is still proportional to $R^{-6}$. Thus, 
to a good approximation, we have 
\begin{align}\label{M63S1Smedium}
{\calW}_{3S;1S}^\mix(R) \approx 
{\overline {\calW}}_{3S;1S}^\mix(R) = 
- \frac{3^9\times 23}{2^{15}}\, E_h \left( \frac{a_0}{R} \right)^6 
\end{align}
in the intermediate distance range.
The mixed pole term is obtained as
\begin{multline}
\label{mixpole3S}
\mathcal{P}_{3S;1S}^{(\mathrm{mix})}(R)= 
\frac{2^{15}\times \sqrt{3}}{3^2\times 5^{6}} \, \frac{E_h a_0^6}{R^6}\,  
\alpha^{\rm dl}_{3S\underline{1S}}\left(\frac{5\,E_h}{72\hbar}\right)
\\
\times\left\{ \cos \left(\frac{5 \alpha R}{36\, a_0}\right)
\left[ 3-5 \left(\frac{5 \alpha R}{72\, a_0}\right)^2 \right.
+ \left(\frac{5 \alpha R}{72\, a_0}\right)^4\right] 
\\
+ \frac{5 \alpha R}{36\, a_0} \, \sin \left(\frac{5 \alpha R}{36\, a_0}\right)
\left[ 3- \left(\frac{5 \alpha R}{72\, a_0}\right)^2\right]\Bigg\},
\end{multline}
where $\alpha^{\rm dl}_{3S\underline{1S}}$ 
represents the dimensionless mixed $\alpha_{3S\underline{1S}}$
polarizability, defined according to Eq.~\eqref{polb}.

%
% Table 3: Fmn, (dimensionless direct dipole matrix element )
%
%\begingroup
%\squeezetable
\begin{table*}[t!]
\begin{center}
\begin{minipage}{0.9\linewidth}
\begin{center}
\caption{\label{table3}
Dimensionless dipole matrix elements $F_{mn}$. For given $m$, dipole
matrix elements decrease with $n$. For given $n$, 
they grow with $m$. Most of the matrix elements  are
expressed in terms of their prime factors. Some of them are given as the
approximate real numbers in order to save space in the table. } 
%{\scriptsize
\begin{tabular}{cccccc}
\hline
\hline
\rule[-2mm]{0mm}{5.5mm}
\diagbox{$n$}{$m$} & 
\multicolumn{1}{c}{2}&
\multicolumn{1}{c}{3} & 
\multicolumn{1}{c}{4} &
\multicolumn{1}{c}{5}&
\multicolumn{1}{c}{6}\\
\hline
\rule[-2mm]{0mm}{5.5mm}
3  &  $\tfrac{2^{15}\,3^{8}}{5^{12}}$ &--  &  -- & --   &  --\\
\rule[-2mm]{0mm}{5.5mm}
4  &   $\tfrac{2^{21}}{3^{15}}$ & $\tfrac{2^{29}\,3^7\,13^2}{7^{16}}$  & -- & --   &  --   \\
\rule[-2mm]{0mm}{5.5mm}
5    &    $\tfrac{2^{15}\,3^3\,5^9}{7^{16}}$   &   $\tfrac{3^7\,5^9\,11^2}{2^{39}}$     &    $\tfrac{2^{22}\,5^{10}\,1447^2}{3^{39}}$ &  -- & --\\
\rule[-2mm]{0mm}{5.5mm}
6    &   $\tfrac{3^8}{2^{18}}$    &   $\tfrac{2^{20}\,11^2}{3^{18}}$    &   $\tfrac{2^{19}\,3^{8}\,3329^2}{5^{23}}$    &  $\tfrac{2^{20}\,3^{8}\,5^7\,67^2\,14969^2}{11^{24}}$  &  --\\
\rule[-2mm]{0mm}{5.5mm}
7   &   $\tfrac{2^{15}\,5^8\,7^9}{3^{41}}$    &  $\tfrac{2^3\,3^7\,7^9\,23^2}{5^{22}}$  &  $\tfrac{5\times2^{22}\,3^{5}\,7^9\,31^2\,233^2}{11^{24}}$   &   $\tfrac{5^{7}\,7^9\,8513^2}{2^{25}\,3^{25}}$    &   $\tfrac{5\times2^{15}\,3^{7}\,7^{10}\,1289^2\,104347^2}{13^{28}}$ \\
\rule[-2mm]{0mm}{5.5mm}
8    &    $\tfrac{2^{30}\,3^9}{5^{22}}$    &   $\tfrac{2^{38}\,3^7\,5^8\,61^2}{11^{24}}$   &   $\tfrac{5\times 2^{25}\,17^2}{3^{23}}$   &  $\tfrac{2^{38}\,3^3\,5^7\,2549^2\,3323^2}{13^{28}}$   &   $\tfrac{5\,2^{30}\,3^{7}\,1051^2\,14327^2}{7^{31}}$  \\
\rule[-2mm]{0mm}{5.5mm}
9   &   $\tfrac{2^{15}\,3^{17}\,7^{12}}{11^{24}}$    &   $\tfrac{3^{11}\,13^2}{2^{29}}$    &   $\tfrac{2^{22}\,3^{17}\,5^7\,17^2\,127^2\,151^2}{13^{28}}$ &  $\tfrac{2^{7}\,3^{17}\,5^7\,13^2\,31^2\,367^2}{7^{30}}$    &   $\tfrac{7 \times 2^{15}\,3^{11}\,179^2\,4451^2}{5^{31}}$   \\
\rule[-2mm]{0mm}{5.5mm}
10   &    $\tfrac{2^{14}\,5^{9}}{3^{27}}$    &    $\tfrac{2^{20}\,3^{7}\,5^9\,7^{12}\,97^2}{13^{28}}$    &   $\tfrac{2^{19}\,3^{13}\,5^{10}\, 709^2}{7^{30}}$    &    $\tfrac{2^{20}\,5^{2}\,73^2\,167^2}{3^{33}}$    &   $\tfrac{3^{7}\,5^{10}\,7^3\,10151^2}{2^{68}}$ \\
\rule[-2mm]{0mm}{5.5mm}
11   &   $\tfrac{2^{15}\,3^{31}\,11^9}{13^{28}}$    &    $\tfrac{2^{25}\,3^{7}\,11^9\,59^2}{7^{30}}$    &   $\tfrac{2^{22}\,7^{12}\,11^{9}\, 254083^2}{3^{31}\,5^{33}}$    &   $\tfrac{3^{11}\,5^{7}\,11^9\,368939^2}{2^{103}}$    &     $\tfrac{7 \times 2^{15}\,3^7\,5^{9}\,11^9\,23^2\,4391^2\,109139^2}{17^{36}}$    \\
\rule[-2mm]{0mm}{5.5mm}
12    &    $\tfrac{2^{21}\,3^{8}\,5^{18}}{7^{30}}$    &    $\tfrac{2^{29}\,3^{18}\,47^2}{5^{32}}$     &   $\tfrac{5\times3^{8}\,107^2}{2^{32}}$   &   $\tfrac{2^{29}\,3^{8}\,5^7\, 7^{12}\,123783047^2}{17^{36}}$     &     $\tfrac{5\times7\times 2^{21}\,23^2\,1847^2}{3^{36}}$   \\
\end{tabular}  %}
%%%%%%%%%
%{\scriptsize
\begin{tabular}{rrrrrr}
% \begin{tabular}{cccccc}
\hline
\hline
\rule[-2mm]{0mm}{5.5mm}
\diagbox{$n$}{$m$} & 
\multicolumn{1}{c}{7}&
\multicolumn{1}{c}{8}&
\multicolumn{1}{c}{9}&
\multicolumn{1}{c}{10}&
\multicolumn{1}{c}{11}\\
\hline
\rule[-2mm]{0mm}{5.5mm}
8  & 
\multicolumn{1}{c}{$\tfrac{2^{39}\, 7^{7} \, 61^2\, 7243^2 \, 63689^2}{3^{29}\, 5^{32}}$} &      --      &     --    &    --     &     -- \\
\rule[-2mm]{0mm}{5.5mm}
\rule[-2mm]{0mm}{5.5mm}
9  & \multicolumn{1}{c}{$\tfrac{3^{17}\, 7^{7} \, 83^2 \, 58991011^2}{2^{106}}$} 
&       389.637946068792    &     --    &    --    &     -- \\
\rule[-2mm]{0mm}{5.5mm}
  10  &  10.168685073471  &       53.715308427324    &      632.560631964481    &     --  &     --\\
\rule[-2mm]{0mm}{5.5mm}
  11  &  4.646650394408   &     17.168247963601     &      86.059151181197   &       974.236212439363   &       --\\
\rule[-2mm]{0mm}{5.5mm}
  12 &  2.559407028125  &     7.772807970611    &     27.217069526272  &     131.078342047212  &      1438.320654775515 \\
\hline 
\hline
\end{tabular}  %}
\end{center}
\end{minipage}
\end{center}
\end{table*}
%\endgroup

%
% Table 4
%
%\begingroup
%\squeezetable
\begin{table*}[!thbp]
\begin{center}
\begin{minipage}{0.9\linewidth}
\begin{center}
\caption{\label{table4}
Numerical values of the dimensionless polarizabilities $G_{mn}$.  The
polarizability follows the following trends: It increases with 
the principal quantum number $n$ of the reference state;
for given $n$, the polarizability decreases with $m$ and approaches the
ground state  static value $9/2$ in the limit $m \to \infty$.}
%{\scriptsize
\begin{tabular}{ccccccccccc}
\hline
\hline
\rule[-2mm]{0mm}{5.5mm}
\diagbox{$n$}{$m$} & 
\multicolumn{1}{c}{2}&
\multicolumn{1}{c}{3} & 
\multicolumn{1}{c}{4} &
\multicolumn{1}{c}{5}&
\multicolumn{1}{c}{6}&
\multicolumn{1}{c}{7}&
\multicolumn{1}{c}{8} & 
\multicolumn{1}{c}{9} &
\multicolumn{1}{c}{10}&
\multicolumn{1}{c}{11}\\
\hline
\rule[-2mm]{0mm}{5.5mm}
3  & 4.632\,34  & --  &  -- & --   &   --  &  -- & --& --& --& --\\
\rule[-2mm]{0mm}{5.5mm}
4  & 4.747\,78 & 4.515\,76 &  -- & --   &   --  &  -- & --& --& --& --\\
\rule[-2mm]{0mm}{5.5mm}
5  & 4.815\,70 & 4.533\,88 &  4.503\,37 & --   &   --  &  -- & --& --& --& --\\
\rule[-2mm]{0mm}{5.5mm}
6  & 4.856\,82 & 4.546\,68 &  4.508\,03 & 4.500\,99  &   --  &  -- & --& --& --& --\\
\rule[-2mm]{0mm}{5.5mm}
7  & 4.883\,15 & 4.555\,41 &  4.511\,81 & 4.502\,55 &   4.500\,36 &  -- & --& --& --& --\\
\rule[-2mm]{0mm}{5.5mm}
8  & 4.900\,86  & 4.561\,50 & 4.514\,65 & 4.503\,95  &   4.500\,98 & 4.500\,15 & --& --& --& --\\
\rule[-2mm]{0mm}{5.5mm}
9  & 4.913\,31 & 4.565\,87 & 4.516\,79 & 4.505\,09 &   4.501\,58 & 4.500\,43 & 4.500\,07& --& --& --\\
\rule[-2mm]{0mm}{5.5mm}
10 & 4.922\,36 & 4.569\,09 & 4.518\,40 & 4.505\,99 &   4.502\,10 & 4.500\,72 & 4.500\,21 & 4.500\,04 & --& --\\
\rule[-2mm]{0mm}{5.5mm}
11 & 4.929\,13 & 4.571\,52 & 4.519\,64 & 4.506\,70 &    4.502\,53 &  4.500\,98 & 4.500\,36 & 4.500\,11 & 4.500\,02& --\\
\rule[-2mm]{0mm}{5.5mm}
12 & 4.934\,34 & 4.573\,40  & 4.520\,61  & 4.507\,27  &    4.502\,89 & 4.501\,21& 4.500\,50 & 4.500\,19 & 4.500\,06 & 4.500\,01 \\
\hline
\hline
\end{tabular}  %}
\end{center}
\end{minipage}
\end{center}
\end{table*}
%\endgroup

%
% Table 5: Hmn, dimensionless mixing dipole matrix elements
%
%\begingroup
%\squeezetable
\begin{table*}[!thbp]
\begin{center}
\begin{minipage}{0.9\linewidth}
\begin{center}
\caption{\label{table5}
Dimensionless dipole matrix elements $H_{mn}$. For given $m$, the dipole
matrix elements decrease with the 
reference state quantum number $n$.  Most of the matrix elements  are
presented in terms of their prime factor decompositions. Some of them are given as 
approximate real numbers in order to save some space in the table.} 
%{\scriptsize
\begin{tabular}{cccccc}
\hline
\hline
\rule[-2mm]{0mm}{5.5mm}
\diagbox{$n$}{$m$} & 
\multicolumn{1}{c}{2}&
\multicolumn{1}{c}{3} & 
\multicolumn{1}{c}{4} &
\multicolumn{1}{c}{5}&
\multicolumn{1}{c}{6}\\
\hline
\rule[-2mm]{0mm}{5.5mm}
3  &   $\tfrac{2^{15}}{5^{6}\,3^{1/2}}$& --  &  -- & --   &   --\\
\rule[-2mm]{0mm}{5.5mm}
4  &    $\tfrac{2^{18}}{3^{12}}$  &  $\tfrac{13\times2^{8}\,3^7}{7^{8}}$   &  --      &   --     &   --\\
\rule[-2mm]{0mm}{5.5mm}
5    &   $\tfrac{2^{15}\,5^{9/2}}{3^3\,7^{8}}$   &  $\tfrac{11\times 3^7\,5^{9/2}}{2^{26}}$     &   $\tfrac{1447\times2^{22}}{3^{18}\,5^{3/2}\,}$   &    --    &  --\\
\rule[-2mm]{0mm}{5.5mm}
6    &   $\tfrac{1}{2^{3/2}\,3^{1/2}}$    &  $\tfrac{11\times 2^{7/2}}{3^{11/2}}$    &  $\tfrac{3329 \times 2^{41/2}\,3^{11/2}}{5^{18}}$    &  $\tfrac{67\times14969\times2^{23/2}\,5^7}{3^{7/2}\,11^{12}}$    &   --\\
\rule[-2mm]{0mm}{5.5mm}
7    &   $\tfrac{2^{15}\,5^4\,7^{9/2}}{3^{25}}$    &  $\tfrac{23\times 3^7\,7^{9/2}}{2^{5}\,5^{11}}$    &  $\tfrac{31\times 233\times 2^{22}\,3^{4}\,7^{9/2}}{5^6\,11^{12}}$   &  $\tfrac{8513\times 5^{7}\,7^{9/2} }{2^{11}\,3^{20}}$     &   $\tfrac{1289\times104347\times 2^{15}\,3^{7}\,5^4}{7^{7/2}\,13^{14}}$ \\
\rule[-2mm]{0mm}{5.5mm}
8    &   $\tfrac{2^{45/2}}{5^{11}}$    &   $\tfrac{61\times 2^{25/2}\,3^7\,5^4}{11^{12}}$    &   $\tfrac{17\times 2^{47/2}}{3^{10}\,5^6}$   &   $\tfrac{2549\times 3323\times2^{41/2}\,5^7}{3^6\,13^{14}}$     &   $\tfrac{1051\times 14327\times 2^{45/2}\,3^{7}\,5^4}{7^{24}}$  \\%
\rule[-2mm]{0mm}{5.5mm}
9    &  $\tfrac{2^{15}\,3^{4}\,7^{6}}{11^{12}}$    &  $\tfrac{13\times 3^{9}}{2^{21}}$    &  $\tfrac{ 17\times127\times151 \times 2^{22}\,3^{10}}{5^3\,13^{14}}$   &   $\tfrac{13\times 31\times 367\times 2^{5}\,3 5^7}{7^{15}}$     &   $\tfrac{179\times4451\times 2^{15}\,3^{9}}{5^{12}\,7^8}$   \\%
\rule[-2mm]{0mm}{5.5mm}
10    &   $\tfrac{2^{29/2}\,5^{9/2}}{3^{18}}$    &   $\tfrac{97\times 2^{7/2}\,3^{7}\,5^{9/2}\,7^{6}}{13^{14}}$    &  $\tfrac{709\times2^{41/2}\,3^{8} }{5^{3/2}\,7^{15}}$    &   $\tfrac{73 \times 167\times 2^{23/2}\,5^{9/2}}{3^{24}}$     &   $\tfrac{10151\times 3^{7}\,5^{17/2}}{2^{53/2}\,7^7}$ \\%
\rule[-2mm]{0mm}{5.5mm}
11    &   $\tfrac{2^{15}\,3^{11}\,11^{9/2}}{13^{14}}$    &   $\tfrac{59\times 2^{6}\,3^{7}\,11^{9/2}}{7^{15}}$     &   $\tfrac{254083\times 2^{22}\,7^{6}\,11^{9/2} }{3^{14}\,5^{23}}$    &  $\tfrac{368939\times 5^{7}\,11^{9/2} }{2^{50}\,3^2}$     &   $\tfrac{23\times 4391\times 109139\times2^{15}\,3^7\,5^{8}\,11^{9/2}}{7^8\,17^{18}}$  \\%
\rule[-2mm]{0mm}{5.5mm}
12    &   $\tfrac{2^{18}\,5^{9}}{3^{1/2}\,7^{15}}$    &   $\tfrac{47\times 2^{8}\,3^{25/2} }{5^{16}}$     &   $\tfrac{107\times 3^{11/2}}{2^{5}\,5^6}$   & $\tfrac{123783047\times 2^{16}\,5^7\, 7^{6}}{3^{7/2}\,17^{18}}$     &    $\tfrac{23 \times 1847 \times 2^{18}\,5^4}{3^{29/2}\,7^8}$   \\%
\end{tabular}  %}
%%%%%%%%%
%{\scriptsize
\begin{tabular}{cccccc}
\hline
\hline
\rule[-2mm]{0mm}{5.5mm}
\diagbox{$n$}{$m$} & 
\multicolumn{1}{c}{7}&
\multicolumn{1}{c}{8}&
\multicolumn{1}{c}{9}&
\multicolumn{1}{c}{10}&
\multicolumn{1}{c}{11}\\
\hline
\rule[-2mm]{0mm}{5.5mm}
8   \ & \          1.818\,659\,585\,095     \ & \       --   \ & \      --   \ & \      --   \ & \       -- \\%
\rule[-2mm]{0mm}{5.5mm}
9   \ & \         0.680\,670\,360\,144     \ & \         1.944\,278\,307\,643     \ & \       --   \ & \      --   \ & \      -- \\%
\rule[-2mm]{0mm}{5.5mm}
10    \ & \          0.387\,187\,449\,739    \ & \        0.721\,899\,672\,793    \ & \        2.063\,501\,367\,283    \ & \       --    \ & \       --\\%
\rule[-2mm]{0mm}{5.5mm}
11   \ & \        0.261\,733\,234\,429   \ & \        0.408\,122\,742\,837     \ & \         0.761\,118\,380\,971      \ & \         2.177\,019\,770\,691     \ & \       --\\%
\rule[-2mm]{0mm}{5.5mm}
12   \ & \          0.194\,248\,892\,826     \ & \        0.274\,610\,271\,380    \ & \         0.428\,030\,151\,135     \ & \         0.798\,538\,222\,524     \ & \        2.285\,469\,435\,260 \\%
\hline 
\hline
\end{tabular}  %}
\end{center}
\end{minipage}
\end{center}
\end{table*}
%\endgroup
%

%
% Limit of Large Distance
%
\subsection{Very large interatomic distance} 
\label{subsec:AwayD}

We are entering the regime
\begin{equation}
\label{regionIII}
R \gg \frac{\hbar c}{{\calL}} \,.
\end{equation}
This range is irrelevant for interactions in the laboratory but
not for interactions relevant to
astrophysical processes~\cite{SwHi2008,ChSu2008}.
Expressed in units of the Hartree energy $E_h$,
the physical values of the Lamb shift and fine structure
energies are~\cite{JeKoLBMoTa2005,hdel,FaPi1971}
\begin{subequations} \label{eq:LFValues}
\begin{align}
{\calL}_3 =& \; 4.78 \times 10^{-8} \, E_h \,,
\\[0.1133ex]
{\calF}_3 =& \; 4.46 \times 10^{-7} \, E_h \approx 10 \, {\calL}_3 \,.
\end{align}
\end{subequations}
The approximation~\eqref{regionIII} is valid in the
region
\begin{equation}
\label{region3}
R \gg \frac{\hbar c}{{\calL}_3} =
\frac{a_0}{\alpha} \frac{E_h}{{\calL}_3} =
2.864 \times 10^9 a_0 =
0.1516 \, {\rm m}  \,.
\end{equation}

For very large interatomic separation $R \to \infty$,
the integrands in Eqs.~\eqref{widetildeWdir}
and~\eqref{overlineWdir} are significantly damped 
by exponential damping in $\omega$.
For large $R$, we may thus carry out the following
approximations in the integrands of the 
\vdw{} energy [Eqs.~\eqref{widetildeWdir} and~\eqref{overlineWdir}]
\begin{equation}
\label{static}
\alpha_{3S}(\omega) \approx \alpha_{3S}(0) \approx \overline\alpha_{3S}(0) \,,
\qquad
\alpha_{1S}(\omega) \approx \alpha_{1S}(0) \,.
\end{equation}
The remaining integral is evaluated as follows,
\begin{align} \label{eq:MJFar}
& {\rm Re} \,  \frac{\hbar}{\pi  c^4} \,
\int\limits_0^\infty {\rm d}\omega \,
{\rm e}^{-2 \omega R/c} \, \frac{\omega^4}{R^2}\,
\left[ 1 + 2 \frac{c}{\omega R}
+ 5 \left( \frac{c}{\omega R} \right)^2 \right.
\nonumber\\[0.1133ex]
& \qquad \left. + 6 \left( \frac{c}{\omega R} \right)^3
+ 3 \left( \frac{c}{\omega R} \right)^4
\right] = \frac{23}{4 \pi} \, \frac{\hbar c}{R^7} \,.
\end{align}
The interaction is known as the retarded Casimir--Polder 
interaction and is proportional to $R^{-7}$.

The dominant contribution to the static polarizability
of the excited $3S$ state comes from the virtual 
$3P_{1/2}$ and $3P_{3/2}$ levels,
\begin{equation} \label{eq:Static2SBar}
\overline \alpha_{3S}(0) 
%= 2 \overline P_{3S}(0) \approx 
%G \, e^2 \, a_0^2 \,  
%\left( \frac{2}{\calF} - \frac{1}{\calL} \right)\,,
= 2 \overline P_{3S}(0) =
36 \, e^2  \,  a_0^2 \,
\left( \frac{2}{{\calF}_3} - \frac{1}{{\calL}_3} \right)\,.
\end{equation}
From Eqs.~\eqref{static1S},~\eqref{eq:MJFar} and 
\eqref{eq:Static2SBar}, 
we find that the large-distance limit {\em of the 
Wick-rotated contribution} to the 
$3S$--$1S$ interaction energy is positive (repulsive),
\begin{equation}
\label{D72S1S}
{\calW}_{3S;1S}^{\left(\mathrm{dir}\right)}\left(R\right) 
\;\;
\mathop{= }^{R \to \infty}
\;\;
\frac{1863}{2 \pi \alpha} E_h \,
\left( \frac{a_0}{R} \right)^7 \,
\left( \frac{E_h}{{\calL}_3}  -2 \frac{E_h}{{\calF}_3} \right) \,.
\end{equation}
This interaction is valid only for very large
interatomic distances given in Eq.~\eqref{region3}.

The dominant term 
in the range~\eqref{regionIII} 
comes from the pole contribution in Eq.~\eqref{Pole3S1SCP}
and reads
\begin{align}
\label{Pole3S1SLSdominant}
\calP_{3S;1S}^{(\mathrm{dir})}&(R)=
- \frac{2^{3}}{5^{8}} \frac{\alpha^4 \, E_h}{\rho^2}
\alpha^{\rm dl}_{1S}\left(\frac{5\,E_h}{72\hbar}\right) \,
\cos\left(\frac{5 \alpha \rho}{36}\right) \,,
\end{align}
where $\rho = R/a_0$.
The pole term falls off as $R^{-2}$
and dominates the interaction energy (see Fig.~\ref{fig1}).
We note the numerical identities
\begin{equation}
\label{Pole3S1SLSdominant_1}
\frac{2^{3}}{5^{8}} = 2.048 \times 10^{-5} \,,
\qquad
\alpha^{\rm dl}_{1S}\left(\frac{5\,E_h}{72\hbar}\right) =
4.63234 \,.
\end{equation}
The coefficient multiplying the leading 
oscillatory $1/R^2$ term given in Eq.~\eqref{Pole3S1SLSdominant}
thus is of order $10^{-4}$; this is in contrast to the 
$D_6$ and $\overline D_6$ coefficients, which are 
of order $10^3$ (in atomic units). The numerical coefficients are thus 
in part responsible for a certain suppression of the 
long-range tail, as evident (in the intermediate region) 
from Fig.~\ref{fig2}.
The same trend is observed for $nD$--$1S$ interactions~\cite{JeAdDe2017prl}.

We should supplement the result for the mixing term
in the very-long-range~\eqref{region3}.
As far as the mixing type contribution to the 
Casimir--Polder term is concerned, 
the degenerate part dominates the nondegenerate one.
One has
\begin{multline}
\label{M73S1S}
{\calW}_{3S;1S}^{\left(\mathrm{mix}\right)}\left(R\right) \approx 
\overline{W}_{3S;1S}^{\left(\mathrm{mix}\right)}\left(R\right)
\\[0.1133ex]
= \frac{3^7\times 23^2}{2^{16}}\frac{E_h}{\pi \alpha}
\left( \frac{a_0}{R} \right)^7
\left( \frac{E_h}{{\calL}_3}  -2 \frac{E_h}{{\calF}_3} \right) \,.
\end{multline}
By contrast, the leading $1/R^2$ 
contribution to the mixing pole term
reads as [see Eq.~\eqref{mixpole3S}]
\begin{align}
\mathcal{P}_{3S;1S}^{(\mathrm{mix})}(R)=& \;
- \frac{2^{3}\times \sqrt{3}}{3^{10} \times 5^2} \, \alpha^4 E_h \,
\left( \frac{a_0}{R} \right)^2 \,   
\nonumber\\
&\times \alpha^{\rm dl}_{3S\underline{1S}}\left(\frac{5\,E_h}{72\hbar}\right) 
\cos\left(\frac{5 \alpha R}{36\,a_0}\right) \,,
\end{align}
and it dominates in the very-long-range limit [Eq.~\eqref{region3}].

%
% States with $4 \leq n \leq 12$
%
\section{States with 
\texorpdfstring{$\maybebm{4 \leq n \leq 12}$}{4 <= n <= 12}}
\label{sec4}

%
% \VDW{} range
%
\subsection{\VDW{} range}

First, we discuss the $nS$--$1S$ interaction, 
with $4 \leq n \leq 12$, in the \vdw{} regime \eqref{vdwRegime},
\begin{equation}
a_0 \ll R \ll \frac{a_0}{\alpha} \,.
\end{equation}
In this range, the interaction is described to 
good approximation by the functional form~\eqref{C6D6M6}.
One should mention that the calculation of 
polarizability-type matrix elements that
generalize Eq.~\eqref{tildeP3S} to states with 
$n \geq 4$ requires the
sophisticated use of contiguous relations for
hypergeometric functions~\cite{Ba1953vol1,Ba1953vol2}.
Eventually, one can bring the matrix elements into 
a form that involves a rational function of the variable
\begin{equation}
\tau_n = \left( 1 + \frac{n^2 \hbar \omega}{\alpha^2 m_e c^2} \right)^{1/2}\,,
\end{equation}
where $n$ is the principal quantum number, 
and a further term where a rational 
function of $\tau_n$ multiplies the hypergeometric function
\begin{equation}
{}_2 F_1\left( 1, -n \tau_n, 
1 - n \tau_n, \frac{(1-\tau_n)^2}{(1+\tau_n)^2} \right) \,.
\end{equation}
Here we describe calculations of the polarization-type 
matrix elements~\eqref{PnStilde} for $nS$ states with principal
quantum numbers as high as n = 12; several thousand terms 
are encountered in intermediate steps of the 
calculations; these are handled with the help of 
computer algebra systems~\cite{Wo1999}.
For the mixed polarizabilities given in Eqs.~\eqref{pola}
and~\eqref{polb}, the calculations are even a little more involved
because the radial wave functions of the 
bra and ket states are different; one may still express
them in terms of a rational function of the $\tau_n$ 
variable, and a hypergeometric function.
Note that lattice methods that lead to a pseudospectrum
of virtual states (see Ref.~\cite{SaOe1989})
cannot be used with good effect for highly excited states,
because of numerical problems associated with the 
modeling of wave functions with many nodes.
These numerical difficulties may be one reason why 
early numerical calculations for 
$C_6(2S;1S)$ coefficients~\cite{Ch1972,DeYo1973}
were never generalized to higher excited $S$ states.
Eventually, for $4 \leq n \leq 12$,
the $D_6$ and $M_6$ coefficients are given in Tables~\ref{table1}
and~\ref{table2} as the generalizations of
Eqs.~\eqref{C62S1Sshortnum} and~\eqref{M62S1Sshortnum},
respectively.
 
%
% Table 6: I_{mn} dimensionless ground state polarizability (mixing)
%
%\begingroup
%\squeezetable
\begin{table*}[th!]
\begin{center}
\begin{minipage}{0.9\linewidth}
\begin{center}
\caption{\label{table6} Numerical values of the dimensionless polarizabilities
$I_{mn}$. In contrast to the dimensionless polarizabilities $G_{mn}$, 
the trend in the numerical data implies lower values of  $I_{mn}$ 
for higher excited reference states.
Also, $I_{mn}$ for given $n$ decreases as the value of $m$ increases.} 
%{\scriptsize
\begin{tabular}{ccccccccccc}
\hline
\hline
\rule[-2mm]{0mm}{5.5mm}
\diagbox{$n$}{$m$} & 
\multicolumn{1}{c}{2}&
\multicolumn{1}{c}{3} & 
\multicolumn{1}{c}{4} &
\multicolumn{1}{c}{5}&
\multicolumn{1}{c}{6}&
\multicolumn{1}{c}{7}&
\multicolumn{1}{c}{8} & 
\multicolumn{1}{c}{9} &
\multicolumn{1}{c}{10}&
\multicolumn{1}{c}{11}\\
\hline
\rule[-2mm]{0mm}{5.5mm}
3  & -2.159\,39 & --  &  -- & --   &   --  &  -- &--&--&--&--\\%
\rule[-2mm]{0mm}{5.5mm}
4  & -1.181\,40 & -1.135\,68  &  -- & --   &   --  &  -- &--&--&--&--\\%
\rule[-2mm]{0mm}{5.5mm}
5  & -0.782\,85 & -0.749\,72 &  -0.745\,94 & --   &   --  &  -- &--&--&--&--\\%
\rule[-2mm]{0mm}{5.5mm}
6  & -0.571\,51 &-0.546\,45 &  -0.543\,13 & -0.542\,52  &   --  &  -- &--&--&--&--\\%
\rule[-2mm]{0mm}{5.5mm}
7  & -0.442\,47  & -0.422\,73 &  -0.419\,92 & -0.419\,32  &   -0.419\,17 &  -- &--&--&--&--\\%
\rule[-2mm]{0mm}{5.5mm}
8  &-0.356\,43  & -0.340\,36  & -0.337\,98 & -0.337\,43  &   -0.337\,28 &  -0.337\,23  &--&--&--&--\\
\rule[-2mm]{0mm}{5.5mm}
9  &-0.295\,46 & -0.282\,06  & -0.280\,02  & -0.279\,53  &   -0.279\,38  & -0.279\,33  &-0.279\,32&--&--&--\\
\rule[-2mm]{0mm}{5.5mm}
10 & -0.250\,31 & -0.238\,91   & -0.237\,15 & -0.236\,71 &   -0.236\,57  & -0.236\,52  & -0.236\,50  & -0.236\,50 &--&--\\
\rule[-2mm]{0mm}{5.5mm}
11 & -0.215\,54  & -0.205\,87 & -0.204\,32  & -0.203\,93  &     -0.203\,80  &  -0.203\,76 &  -0.203\,74 &  -0.203\,73  &-0.203\,73 &--\\
\rule[-2mm]{0mm}{5.5mm}
12 & -0.188\,50 & -0.179\,87  & -0.178\,50  & -0.178\,15  &    -0.178\,04 & -0.177\,99  & -0.177\,98 & -0.177\,97  & -0.177\,96  &-0.177\,96\\
\hline
\hline
\end{tabular}  %}
\end{center}
\end{minipage}
\end{center}
\end{table*}
%\endgroup

%
% Table 7
%
\begin{table}[t!]
\begin{center}
\caption{\label{table7} 
Numerical values of the degenerate contributions to the direct $D_7$
and mixed $M_7$ Casimir-Polder coefficients for two-atom systems. The
coefficients are given in units of
$\tfrac{1}{\alpha\pi}E_h^2\left(-\tfrac{1}{\mathcal{L}_n} +
\tfrac{2}{\mathcal{F}_n}\right)a_0^7$
(which is a negative quantity), explaining why the 
overall interaction term is repulsive.}
\begin{tabular}{lcr}
\hline
\hline
\rule[-2mm]{0mm}{5.5mm}
System & $D_7$ & $M_7$ \\
\hline
\rule[-2mm]{0mm}{5.5mm}
$3S$--$1S$\ &\ $\tfrac{1\,863}{2}$\ &\ 17.653\,244\,019\\
\rule[-2mm]{0mm}{5.5mm}
$4S$--$1S$\ &\ $3105$\ &\ 10.241\,950\,813\\
\rule[-2mm]{0mm}{5.5mm}
$5S$--$1S$\ &\ $\tfrac{15\,525}{2}$\ &\ 7.304\,926\,315\\
\rule[-2mm]{0mm}{5.5mm}
$6S$--$1S$\ &\ $\tfrac{65\,205}{4}$\ &\ 5.725\,196\,938\\
\rule[-2mm]{0mm}{5.5mm}
$7S$--$1S$\ &\ 30\,429\ &\ 4.730\,674\,062\\
\rule[-2mm]{0mm}{5.5mm}
$8S$--$1S$\ &\ 52\,164\ &\ 4.042\,550\,036\\
\rule[-2mm]{0mm}{5.5mm}
$9S$--$1S$\ &\ 83\,835\ &\  3.535\,767\,981\\
\rule[-2mm]{0mm}{5.5mm}
$10S$--$1S$\ &\ $\tfrac{512\,325}{4}$\ &\ 3.145\,707\,517\\
\rule[-2mm]{0mm}{5.5mm}
$11S$--$1S$\ &\ $\tfrac{375\,705}{2}$\ &\ 2.835\,483\,977\\
\rule[-2mm]{0mm}{5.5mm}
$12S$--$1S$\ &\ 266\,409 \ &\ 2.582\,435\,832\\
\hline
\hline
\end{tabular}
\end{center}
\end{table}

%
% table 8 
%
\begin{table}[t!]
\begin{center}
\caption{\label{table8} Numerical values of the long-range interaction
frequency shift  in the $(3S;1S)$ system. The ${\calW}_{3S;1S}(R)$ is the
Wick-rotated type frequency shift and the $\mathcal{P}_{3S;1S}(R)$ is the pole
type frequency shift. The $\pm$ sign corresponds to the $\pm$ sign in the
$\left( |1S\rangle |3S\rangle \pm |3S\rangle |1S\rangle\right)$ superposition.
For small separation, the Wick-rotated frequency is dominant, however, for 
large separation, the pole term dominates.}
\begin{tabular}{lll}
\hline
\hline
\rule[-2mm]{0mm}{5.5mm}
R [$\mbox{\AA}$] &
\multicolumn{1}{c}{${\calW}_{3S;1S}(R)$ [Hz]} & 
\multicolumn{1}{c}{$\mathcal{P}_{3S;1S}(R)$ [Hz]} \\
\hline
\rule[-2mm]{0mm}{5.5mm}
$20 $    &  $-\left( 2.053\pm 0.019\right) \!\times\!10^{9}$  & $-\left(1.842\mp 1.181\right) \!\times\!10^{7}$\\
\rule[-2mm]{0mm}{5.5mm}
$40 $    &  $-\left( 3.207\pm 0.029\right) \!\times\!10^{7}$  & $-\left(2.876\mp 1.185\right) \!\times\!10^{5}$\\
\rule[-2mm]{0mm}{5.5mm}
$80 $    &  $-\left( 5.007\pm 0.045\right) \!\times\!10^{5}$  & $-\left(4.505\mp 2.888\right) \!\times\!10^{3}$\\
\rule[-2mm]{0mm}{5.5mm}
$200 $   &  $-\left( 2.045\pm 0.019\right) \!\times\!10^{3}$  & $-\left(1.865\mp 1.195\right) \!\times\!10^{1}$\\
\rule[-2mm]{0mm}{5.5mm}
$400 $   &  $-\left( 3.174\pm 0.030\right) \!\times\!10^{1}$  & $-\left(3.035\mp 1.946\right) \!\times\!10^{-1}$\\
\rule[-2mm]{0mm}{5.5mm}
$800 $   &  $-\left( 4.890\pm 0.050\right) \!\times\!10^{-1}$ & $-\left(5.561\mp 3.565\right) \!\times\!10^{-3}$\\
\rule[-2mm]{0mm}{5.5mm}
$2\,000$ &  $-\left( 2.328\pm 0.012\right) \!\times\!10^{-3}$ & $-\left(9.460\mp 2.021\right) \!\times\!10^{-4}$\\
\rule[-2mm]{0mm}{5.5mm}
$20\,000$&  $-\left( 1.714\pm 0.029\right)\!\times\!10^{-9}$  & $-\left(1.835\mp 0.392\right) \!\times\!10^{-7}$\\
\rule[-2mm]{0mm}{5.5mm}
$200\,000$ & $-\left( 1.653\pm 0.031\right)\!\times\!10^{-15}$ & $-\left(4.032\mp 0.862\right) \!\times\!10^{-9}$\\
\hline
\hline
\end{tabular}
\end{center}
\end{table}

%
% table 9 
%
\begin{table}[t!]
\begin{center}
\caption{\label{table9} Numerical values of the Wick-rotated type
${\calW}_{4S;1S}(R)$ and the pole type $\mathcal{P}_{4S;1S}(R)$  long-range
interaction frequency shift  in the $(4S;1S)$ system. The $\pm$ sign
corresponds to the $\pm$ sign in the $\left( |1S\rangle |4S\rangle \pm
|4S\rangle |1S\rangle\right)$ superposition.}
\begin{tabular}{lll}
\hline
\hline
\rule[-2mm]{0mm}{5.5mm}
R ($ \mbox{\AA}$ ) &
\multicolumn{1}{c}{${\calW}_{4S;1S}(R)$ [Hz]} & 
\multicolumn{1}{c}{$\mathcal{P}_{4S;1S}(R)$ [Hz]} \\
\hline
\rule[-2mm]{0mm}{5.5mm}
$20 $    & $-\left(6.425\pm 0.011 \right) \!\times\!10^{9}$&$-\left(1.249\mp 0.091\right) \!\times\!10^{8}$\\
\rule[-2mm]{0mm}{5.5mm}
$40 $    & $-\left(1.004\pm 0.002 \right) \!\times\!10^{8}$&$-\left(1.951\mp 0.146\right) \!\times\!10^{6}$\\
\rule[-2mm]{0mm}{5.5mm}
$80 $    & $-\left(1.568\pm 0.003 \right) \!\times\!10^{6}$&$-\left(3.048\mp 0.229\right) \!\times\!10^{4}$\\
\rule[-2mm]{0mm}{5.5mm}
$200 $   & $-\left(6.416\pm 0.011 \right) \!\times\!10^{3}$&$-\left(1.246\mp 0.094\right) \!\times\!10^{2}$\\
\rule[-2mm]{0mm}{5.5mm}
$400 $   & $-\left(1.000\pm 0.002 \right) \!\times\!10^{2}$&$-\left(1.936\mp 0.151\right) \!\times\!10^{0}$\\
\rule[-2mm]{0mm}{5.5mm}
$800 $   & $-\left(1.554\pm 0.003\right) \!\times\!10^{0}$&$-\left(2.962\mp 0.256\right) \!\times\!10^{-2}$\\
\rule[-2mm]{0mm}{5.5mm}
$2\,000$ & $-\left(8.851\pm 0.003 \right) \!\times\!10^{-3}$&$-\left(5.984\mp 0.462\right) \!\times\!10^{-3}$\\
\rule[-2mm]{0mm}{5.5mm}
$20\,000$& $-\left(5.822\pm0.016\right) \!\times\!10^{-9}$&$-\left(1.152\mp 0.994\right) \!\times\!10^{-6}$\\
\rule[-2mm]{0mm}{5.5mm}
$200\,000$& $-\left(5.519\pm0.018\right) \!\times\!10^{-15}$&$-\left(1.658\mp 1.321\right) \!\times\!10^{-8}$\\
\hline
\hline
\end{tabular}
\end{center}
\end{table}
 
%
% Intermediate distance
%
\subsection{Intermediate distance}

We discuss the intermediate distance range
\begin{equation}
\frac{a_0}{\alpha} \ll R \ll \frac{\hbar c}{\mathcal{L}} \,.
\end{equation}
In Table~\ref{table2},
we generalize the result~(\ref{C62S1Smedium}) and~(\ref{M63S1Smedium})
to higher excited $nS$ states. 
The nonretarded $1/R^6$ tail of the direct term has the 
functional form
\begin{equation}
\label{overlineD6nS1Smedium}
{\calW}_{nS;1S}^{\left(\mathrm{dir}\right)}\left(R\right) = 
-\frac{\overline{D}_6\left(nS;1S\right)}{R^6}\,,
\end{equation}
mainly due to the contribution from the degenerate
$nP$ states.
The degenerate $D_6$ coefficient can be brought into
the general form
\begin{equation}
\label{eq:D6BarPattern}
\overline D_6(nS;1S) = \frac{81}{8} n^2 \, (n^2 - 1) \, E_h \, a_0^6 
\end{equation}
for $n \geq 2$. The leading contribution to the mixing term is
\begin{equation}
\label{overlineM6nS1Smedium}
{\calW}_{nS;1S}^{\left(\mathrm{mix}\right)}\left(R\right) = 
-\frac{\overline{M}_6\left(nS;1S\right)}{R^6}\,,
\end{equation}
again due to the contribution from the degenerate
$nP$ states.
In the intermediate range, the Wick-rotated term of order $1/R^6$
competes with the pole term given in Eq.~\eqref{DirectPoleGeneral},
due to lower-lying $mP$ states.
We express the latter as follows,
{\allowdisplaybreaks
\begin{multline}
\label{pdir}
\calP^\dir(R) =
-\frac{2 E_h}{3\rho ^6} \sum_{m = 2}^{n-1} 
\frac{\langle nS | \vec d | mP \rangle \cdot
\langle mP | \vec d | nS \rangle}{e^2 \, a_0^2} \,
\\[0.1133ex]
\times \alpha^{\rm dl}_{1S}\left( \frac{d_{mn} \, \alpha \, E_h}{2 \hbar} \right)
\\[0.1133ex]
\times
\left\{\cos\left( d_{mn} \,  \alpha \, \rho \right)  \,
\left( 3 - 5 \left( \frac{d_{mn} \,  \alpha \, \rho}{2} \right)^2 +
\left( \frac{d_{mn} \,  \alpha \, \rho}{2} \right)^4 \right)
\right.
\\[0.1133ex]
\left. + d_{mn} \,  \alpha \, \rho \,
\sin\left( d_{mn} \,  \alpha \, \rho \right)
\left( 3 - \left( \frac{d_{mn} \, \alpha \, \rho}{2} \right)^2 
\right) \right\}\,,
\end{multline}
}
where $\rho = R/a_0$. We have used the identities
\begin{equation}
\frac{E_{mn}}{\hbar} =
-\frac{d_{mn} \, \alpha \, E_h}{2 \hbar} \,,
\qquad
\frac{2 E_{mn} R}{\hbar c} =
-d_{mn} \,  \alpha \, \rho \,,
\end{equation}
where
\begin{equation}
\label{dmn}
d_{mn} \equiv \frac{1}{m^2} - \frac{1}{n^2}  \,.
\end{equation}
Values for the (dimensionless) dipole matrix elements
\begin{equation}
F_{mn} = \frac{\langle nS | \vec d | mP \rangle \cdot
\langle mP | \vec d | nS \rangle}{e^2 a_0^2} \, 
\quad
m < n \,, 
\quad
n \leq 12 \,,
\end{equation}
are given in Table~\ref{table3}.
While it is possible to give a semi-analytic expression
for the matrix elements
(see the Appendix and Ref.~\cite{Go1929aop}), these
are quite complicated. It is instructive to have an
explicit reference to the absolute magnitude of the coefficients;
hence, we include Table~\ref{table3}.
The (dimensionless) polarizabilities
\begin{equation}
G_{mn} = 
\alpha^{\rm dl}_{1S}\left( \frac{d_{mn} \, \alpha \, E_h}{2 \hbar} \right)
\quad
m < n \,,
\quad
n \leq 12 \,,
\end{equation}
are given in Table~\ref{table4},
for all states relevant to the current investigation.
The real part of the mixing pole term  
has been given in Eq.~\eqref{Pmix},
\begin{multline}
\calP^\mix(R) =
-\frac{2 E_h}{3 \rho^6} \sum_{m < n} 
\frac{\langle nS | \vec d | mP \rangle \cdot
\langle mP | \vec d | 1S \rangle }{e^2 a_0^2}
\\[0.1133ex]
\times 
\alpha^{\rm dl}_{nS\underline{1S}}\left( \frac{d_{mn} \, \alpha \, E_h}{2 \hbar} \right)
\\[0.1133ex]
\times
\left\{\cos\left( d_{mn} \alpha \rho \right)  \,
\left( 3 - 5 \left( \frac{d_{mn} \alpha \rho}{2} \right)^2 +
\left( \frac{d_{mn} \alpha \rho}{2} \right)^4 \right)
\right.
\\[0.1133ex]
\left.+ d_{mn} \alpha \rho \, \sin\left( d_{mn} \alpha \rho \right) \,
\left( 3 - \left( \frac{d_{mn} \alpha \rho}{2} \right)^2 \right) \right\}\,.
\end{multline}
Numerical values for the (dimensionless) dipole matrix elements
\begin{equation}
H_{mn} = 
\frac{\langle nS | \vec d | mP \rangle \cdot
\langle mP | \vec d | 1S \rangle }{e^2 a_0^2} \,,
\quad
m < n \,,
\quad
n \leq 12 \,,
\end{equation}
and the (dimensionless) polarizabilities
\begin{equation}
I_{mn} = 
\alpha^{\rm dl}_{nS\underline{1S}}\left( \frac{d_{mn} \, \alpha \, E_h}{2 \hbar} \right)
\quad
m < n \,,
\quad
n \leq 12 \,,
\end{equation}
are given in Tables~\ref{table5} and~\ref{table6}, respectively.
For a discussion of the evaluation of the $H_{mn}$,
see the Appendix.
As the principal quantum number $n$ of the excited state of
the hydrogen atom interacting with the ground state increases,
it takes longer and longer for the pole term to 
finally assume dominance over the Wick-rotated term 
(see Figs.~\ref{fig3},~\ref{fig4} and~\ref{fig5}).

A few words on the precise formulation of the
intermediate distance range are perhaps in order.
Namely, in principle, one might argue that the intermediate range should be bounded
from above by $\hbar c/\mathcal{F}_n$,
instead of $\hbar c/\mathcal{L}_n$, as the former quantity is
smaller than the latter.
In the rather narrow window where
$\hbar c/\mathcal{F}_n < R < \hbar c/\mathcal{L}_n$,
transitions between $nS$ and $nP_{3/2}$
states are suppressed  by retardation
while those between $nS$ and $nP_{1/2}$ states are not.
We do not enter the details of this regime due to
its narrow character, which would make it difficult
to reliably clarify the asymptotic behavior of the interaction energy.
Mathematically speaking, the inequality $R \ll \hbar c/\mathcal{L}_n$ implies
$R \ll \hbar c/\mathcal{F}_n$ because ${\calF}_n$ and ${\calL}_n$ are
apart by only a single order-of-magnitude [see Eq.~\eqref{eq:LFValues}].
If desired, then
the regime $\hbar c/\mathcal{F}_n < R < \hbar c/\mathcal{L}_n$
could only be accessed reliably by a numerical
calculation.

\begin{figure}[th]
\begin{center}
\begin{minipage}{0.91\linewidth}
\begin{center}
\includegraphics[width=0.91\linewidth]{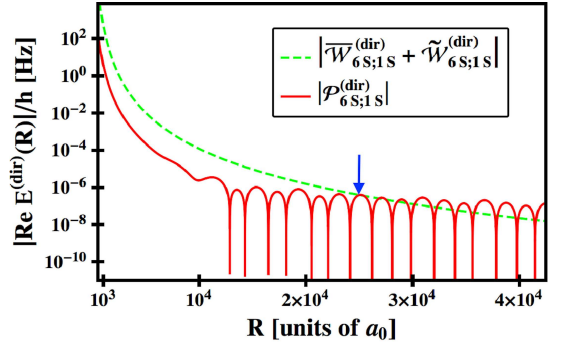}
\end{center}
\end{minipage}
\caption{\label{fig3} Interaction energy in the 
$6S$--$1S$ system as a function of
the interatomic distance, $R$, in the intermediate range. The smooth curve
represents the absolute value of the total Wick-rotated contribution and the
oscillatory curve gives the pole type contribution. The arrow indicates the
minimum  value of R at which the Wick-rotated  and pole terms are equal in
magnitude.}
\end{center}
\end{figure}
  
\begin{figure}[th]
\begin{center}
\begin{minipage}{0.91\linewidth}
\begin{center}
\includegraphics[width=0.91\linewidth]{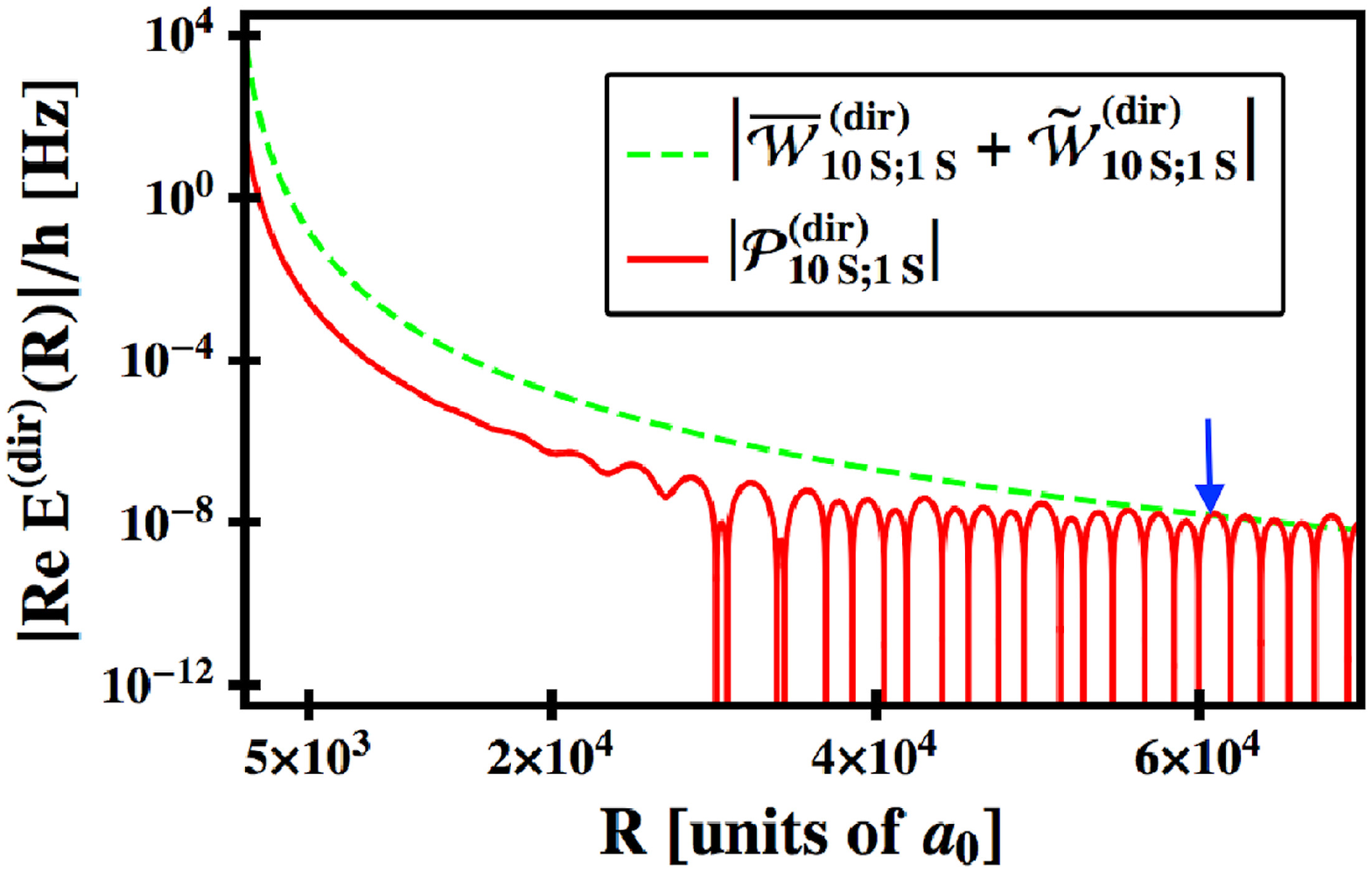}
\end{center}
\end{minipage}
\caption{\label{fig4} 
Interaction energy in the $10S$--$1S$ system 
as a function of the interatomic
distance, $R$, in the intermediate range.The smooth curve represents the
absolute value of the total Wick-rotated contribution and the oscillatory curve
gives the pole type contribution. The arrow indicates the minimum  value of R
at which the Wick-rotated  and pole terms are equal in magnitude.}
\end{center}
\end{figure}  

%%%%%%%
\begin{figure}[th]
\begin{center}
\begin{minipage}{0.91\linewidth}
\begin{center}
\includegraphics[width=0.91\linewidth]{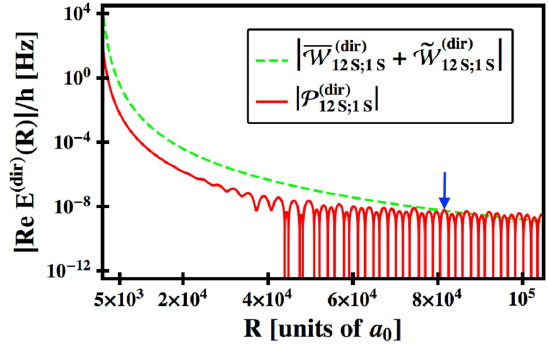}
\end{center}
\end{minipage}
\caption{\label{fig5} Interaction energy in the 
$12S$--$1S$ system as a function of
the interatomic distance, $R$, in the intermediate range. The smooth curve
represents the absolute value of the total Wick-rotated contribution and the
oscillatory curve gives the pole type contribution. The arrow indicates the
minimum  value of R at which the Wick-rotated  and pole terms are equal in
magnitude.}
\end{center}
\end{figure}

%
% Very large distances
%
\subsection{Very large distances}

The regime
\begin{equation}
R \gg \frac{\hbar c}{\calL} 
\end{equation}
is characterized by two competing terms, 
a $1/R^7$ term from the Wick-rotated contribution 
and an oscillatory contribution from the pole term.
In Table~\ref{table7},  we give a generalization of (\ref{D72S1S}) and (\ref{M73S1S})
to higher excited $S$ states,
\begin{equation}
{\calW}_{nS;1S}^{\left(\mathrm{dir}\right)}\left(R\right) =
-\frac{D_7\left(nS;1S\right)}{R^7}\,.
\end{equation}
The $D_7$ coefficients obey the relationship
\begin{equation}
\label{eq:D7BarPattern}
D_7(nS;1S) = \frac{207}{16} n^2 \, (n^2 - 1)\,
\frac{a_0^7\,E_h}{\alpha\pi}\left(-\frac{E_h}{\mathcal{L}_n}+
2\frac{E_h}{\mathcal{F}_n}\right) \,.
\end{equation}
The Wick-rotated contribution to the 
mixing term has the functional form
\begin{equation}
{\calW}_{nS;1S}^{\left(\mathrm{mix}\right)}\left(R\right) =
-\frac{M_7\left(nS;1S\right)}{R^7}\,,
\end{equation}
where we refer to Table~\ref{table7} for the numerical values.
However, the $1/R^7$ tails are suppressed, in the very-long-range
limit, in comparison to the pole terms, which go as $1/R^2$.

In fact, due to the trend in the numerical 
coefficients recorded in Tables~\ref{table3}---\ref{table6},
the dominant contributions from the pole 
terms (direct and mixing term)
comes from virtual $2P$ states and can be expressed as
\begin{multline}
\label{osc1}
\calP^\dir(R) \approx
-\frac{d_{2n}^4 E_h \, \alpha^4}{24 \rho^2} \, F_{2n} \, G_{2n} \,
\cos\left( d_{2n} \,  \alpha \, \rho \right) \,,
\end{multline}
and
\begin{multline}
\label{osc2}
\calP^\mix(R) \approx
-\frac{d_{2n}^4 E_h \, \alpha^4}{24 \rho^2} \, H_{2n} \, I_{2n} \,
\cos\left( d_{2n} \,  \alpha \, \rho \right) \,,
\end{multline}
respectively. Contributions from $mP$ states with 
$3 \leq m \leq n-1$ are numerically, but not parametrically, suppressed.
The quantity $d_{mn}$ was defined in Eq.~\eqref{dmn},
and $\rho$ was defined in the text following Eq.~\eqref{Pole3S1SLSdominant}.

%
% Numerical Examples: Modification to the Hyperfine 
% Structure and Transition Frequencies
%
\section{Numerical Examples}
\label{sec5}

It can be helpful to have 
numerical reference data available for 
the pole term, as well as the Wick-rotated contribution 
to the interaction energy, for sample values
of the interatomic distance. These are 
given in Tables~\ref{table8} and~\ref{table9}.
We concentrate on the $3S$--$1S$ 
and $4S$--$1S$ systems.
One can clearly discern the dominance of the 
pole term in the long-range limit,
and its suppression in the 
\vdw{} range~\eqref{vdWrange}.
Note that both the direct as well as the 
mixing terms are indicated in 
Tables~\ref{table8} and~\ref{table9}.
In entries with $\pm$, the positive sign
refers to the {\em gerade} configuration of the 
wave functions, and the negative sign 
is relevant to the {\em ungerade} 
configuration. The opposite happens for the 
numerical entries involving the $\mp$ sign.

%
% Conclusions
%
\section{Conclusions}
\label{sec6}

We have studied $nS$--$1S$ \vdw{} interactions among
hydrogen atoms in detail, for $n \geq 3$.
In a brief orientation in Sec.~\ref{sec2},
we discuss the nondegenerate Wick-rotated 
contribution ${\widetilde {\calW}}(R)$, 
the degenerate term ${\overline {\calW}}(R)$, 
and the pole term $\calQ(R)$, 
which splits into a real energy shift $\calP(R)$, 
and a width term $\Gamma(R)$.
We treat the $3S$--$1S$ interaction in great
detail in Sec.~\ref{sec3}, before 
generalizing the approach to the 
$nS$--$1S$ case in Sec.~\ref{sec4} ($4 \leq n \leq 12$).
Numerical reference data are given in Sec.~\ref{sec5}.
These numerical data are 
crucial in a reliable determination of 
pressure shifts in high-precision spectroscopy experiments
involving highly excited $S$ states~\cite{dV2002,UdPriv2017}.

We differentiate 
three distance ranges given in Eqs.~\eqref{vdWrange},~\eqref{CPrange1},
and~\eqref{LSrange}, which we recall for convenience:
\begin{subequations}
\begin{align}
\label{three}
\mbox{van der Waals:} & \; \qquad
\frac{\hbar}{\alpha m_e c} \ll R \ll \frac{\hbar}{\alpha^2 m_e c} \,,
\\[0.1133ex]
\mbox{Casimir--Polder:} & \; \qquad
\frac{\hbar}{\alpha^2 m_e c} \ll R \ll \frac{\hbar c}{\mathcal{L}_n} \,,
\\[0.1133ex]
\mbox{Lamb shift:} & \; \qquad
R \gg \frac{\hbar c}{\mathcal{L}_n} \,.
\end{align}
\end{subequations}
In the \vdw{} range, the interatomic interaction is 
described to good accuracy by a functional form $-C_6(A;B)/R^6$,
where $C_6(A;B)=D_6(A;B)\pm M_6(A;B)$ is the \vdw{} coefficient.
The direct coefficient $D_6$
is the sum of a nondegenerate contribution 
${\widetilde D}_6$,
a degenerate contribution $\overline D_6$,
and a pole term $D^\calP_6$.
Analogously, one has
$M_6 = {\widetilde M}_6 + {\overline M}_6 + M^\calP_6$,
where ${\widetilde M}_6$ is the nondegenerate contribution
to the mixing \vdw{} coefficient, while ${\overline M}_6$ and 
$M^\calP_6$ are the degenerate and pole term counterparts.

The main results reported in the 
current investigations can be summarized as follows.
{\em (i)} The \vdw{} coefficients for the
``direct'' and ``mixing'' terms have been 
obtained, for $nS$--$1S$ interactions, in 
Tables~\ref{table1} and~\ref{table2}, on the basis 
of rather involved analytic calculations of 
polarizability-type matrix elements [see Eq.~\eqref{PnStilde}],
with several thousand terms in intermediate 
steps of the calculations; these were handled 
using computer algebra~\cite{Wo1999}.
The data show a surprising trend:
Namely, the $D_6$ and $\overline D_6$
coefficients, as a function of $n$,
are consistent with an $n^4$ asymptotics 
for large $n$ [see also Eq.~\eqref{eq:D6BarPattern}].
By contrast, the mixing coefficients $M_6$ and $\overline M_6$
tend to decrease with $n$.
However, for small $n$ (say, $n=3$), 
in contrast to $nD$--$1S$ interactions~\cite{JeAdDe2017prl}, 
the $M_6$ and ${\overline M}_6$ coefficients 
for $nS$--$1S$ interactions, obtained here,
can be comparatively large and smaller than $D_6$ and $\overline D_6$
by only one order of magnitude.
This situation is completely different for 
$nD$--$1S$ interactions~\cite{JeAdDe2017prl}.

{\em (ii)} We carry out a detailed analysis of the 
oscillatory long-range ``tails'' of the \vdw{}
interaction, for $nS$--$1S$ interactions.
The results obtained for $nS$--$1S$ interactions
indicate that the $1/R^2$ long-range tails are somewhat 
suppressed in comparison to the 
standard $1/R^6$ interaction, due to the smallness
of the overall numerical factors multiplying the 
energy shifts. For example, for $3S$--$1S$ interactions,
one should compare the overall prefactor 
in Eq.~\eqref{Pole3S1SCP}, which is 
\begin{equation}
\frac{2^{15} \times 3^8}{5^{12}} \approx 0.88060 \,,
\end{equation}
with the magnitude of the $D_6$ coefficient, given 
as $D_6 \approx 917.478$ according to Table~\ref{table1}.
We also refer to Eqs.~\eqref{Pole3S1SLSdominant}
and~\eqref{Pole3S1SLSdominant_1}
for the overall prefactors 
multiplying the oscillatory tail, for the $3S$--$1S$ interaction.
For $12S$--$1S$ interactions, the situation is even more extreme:
The leading contribution to the pole term, as far as the 
energy difference $E_{mn}$ is concerned, comes from 
a virtual $2P$ state; the overall coefficient in the pole 
term comes from Table~\ref{table1} as
\begin{equation}
\frac{2^{21} \times 3^8 \times 5^{18}}{7^{30}} \approx 0.00233 \,,
\end{equation}
while the $D_6$ coefficient is as large as $241176$
(see Table~\ref{table1}). 
The difference by several orders-of-magnitude 
between the overall multiplying coefficients
does not originate from a parametric suppression of the 
pole terms, but is exclusively due to the dependence 
of the transition energies and dipole transition 
matrix elements on the quantum numbers of the 
involved states.}
The trend of the coefficients
has the following consequences for the 
physical nature of the interaction: In the intermediate range, 
the non-retarded, quasi-degenerate 
$1/R^6$ contributions
to $\overline D_6$ and ${\overline M}_6$
compete with the oscillatory long-range tail 
of the $1/R^2$ pole term (see Figs.~\ref{fig1}--\ref{fig5}).
As the principal quantum number increases, it takes
longer and longer for the 
pole term to assume dominance over the 
non-retarded tail of the \vdw{} interaction, 
with the latter being given in Eq.~\eqref{eq:D6BarPattern}.

{\em (iii)} The analysis 
presented here also raises interesting further
questions. E.g., for $nS$--$1S$ interactions,
the oscillatory cosine terms,
proportional to $R^{-2}$, eventually dominate in the 
long-range limit [see Eqs.~\eqref{osc1} and~\eqref{osc2}],
and the Casimir-Polder tail of order $R^{-7}$ 
is found to be phenomenologically irrelevant for 
interactions involving higher excited states.
Based on a parametric analysis,
one might think that the $R^{-2}$ oscillatory
tails should also dominate over the 
$R^{-6}$ \vdw{} interactions, in the intermediate 
range of interatomic distances.
However, as evident from 
Figs.~\ref{fig1},~\ref{fig3},~\ref{fig4}, and~\ref{fig5},
the dominance sets in only after 
the absolute magnitude of the energy shift has 
decreased to well below $1\,{\rm Hz}$ in frequency units.
As already stated,
one can attempt to justify this trend based on the 
dependence of the energy differences $d_{mn}$
on the principal quantum numbers.
For example, one has $d_{(n-1)n} \sim n^{-3}$
and the fact that $d_{mn}$ enters the leading $R^{-2}$ contribution
to the pole term in the fourth power
[see Eq.~\eqref{pdir}].
This compensates the growth of the $F_{mn}$ given in 
Table~\ref{table3} with $m$ for given $n$,
and suppresses the contribution from energetically close,
lower-lying virtual states to the pole terms,
for given $n$ of the reference state. 
We also observe the 
decreasing trend in the dipole matrix elements
given in Table~\ref{table3} with $n$, for given $m$.
However, it would be interesting to investigate if there is further, 
deeper reason for the apparent, non-parametric
(there is no factor of the fine-structure constant involved)
suppression of the pole terms, and mixing terms, in long-range interactions
involving higher excited states of simple 
atomic systems. This analysis is left for further study.

%
% Acknowledgments
%
\section*{Acknowledgments}

This project was supported by the National Science Foundation
(Grants PHY--1403973 and PHY--1710856) and by the Missouri Research Board.
The high-precision experiments carried
out at MPQ Garching under the guidance of Professor T.~W.~H\"{a}nsch
have been a major motivation and inspiration for the current theoretical
work. The authors also acknowledge helpful conversations with
A.~Matveev and N.~Kolachevsky.

\appendix

%
% Dipole matrix elements F_{mn} and H_{mn}
%
\section{Dipole matrix elements \\
\texorpdfstring{$\maybebm{F_{mn}}$}{Fmn} and 
\texorpdfstring{$\maybebm{H_{mn}}$}{Hmn}}

We are concerned with the evaluation of dipole matrix 
elements of bound-state Schr\"{o}dinger 
hydrogen wave functions,
\begin{align}
G^{\ell_1 \ell \ell_2}_{n_1nn_2}=
\langle n_1\ell_1m_1 |\vec{d}| n\ell m\rangle\cdot \langle 
n\ell m |\vec{d}| n_2\ell_2m_2\rangle,
\end{align}
where $\vec d = e \, \vec r$ is the dipole operator and 
the dimensionless dipole matrix elements $F_{mn}$ and  $H_{mn}$ are given by 
\begin{align}
F_{mn}= \frac{G^{010}_{nmn}}{e^2 a_0^2} \,, \qquad
\qquad H_{mn}= \frac{G^{010}_{nm1}}{e^2 a_0^2} \,.
\end{align}
The well-known expression for the 
radial function $R_{n\ell}(r)$, 
\begin{align}
\label{radialwavefun}
R_{n\ell}(r)& = 
\sqrt{\frac{(n-\ell-1)!}{(n+\ell)!}} \, 
\frac{2}{n^2a_0^{3/2}} \left(\frac{2 r}{a_0n}\right)^\ell
\nonumber\\
&\times \exp\left(-\frac{r}{a_0n}\right) 
{L}_{n-\ell-1}^{2\ell+1}\left(\frac{2r}{na_0}\right),
\end{align}
allows us to bring the dipole transition matrix element 
into the form
\begin{multline}
\label{4int}
\int_0^\infty \dd r \, r^3 
R_{n'\ell'}(r) R_{n\ell}(r) =  
\sqrt{\frac{(n'-\ell'-1)!}{(n'+\ell')!}} 
\\
\times 
\sqrt{\frac{(n-\ell-1)!}{(n+\ell)!}} 
\frac{a_0}{4 \, n^{\ell + 2}\, n'^{\ell' + 2}}\,
\int_0^\infty \dd x \, 
x^{3+\ell+\ell'} 
\\
\times 
\exp\left[-\frac{x}{2}\left(\frac{1}{n}+\frac{1}{n'}\right)\right] \,
L_{n'-\ell'-1}^{2\ell'+1}\left( \frac{x'}{n'} \right) \,
L_{n-\ell-1}^{2\ell+1}\left( \frac{x}{n} \right) \,,
\end{multline}
where $x = 2r/a_0$.
A result obtained for the radial matrix 
element in Ref.~\cite{Go1929aop} is
reproduced in Eq.~(63.2) of Ref.~\cite{BeSa1957};
the latter appears to benefit from some corrections 
for typographical errors 
that occurred in the original work~\cite{Go1929aop}.
Direct application of Eq.~(63.2) of Ref.~\cite{BeSa1957}
leads to the formula
\begin{subequations} 
\label{resG1}
\begin{align} 
\label{Gn1nn2:ell:1}
G_{n_1nn_2}^{010}=& \; (-1)^{n_1+n_2}
\frac{16\, n^5\, n_1^{5/2}\, n_2^{5/2}
\left(n^2-1\right)e^2 a_0^2}{\left(n-n_1\right)^4
\left(n-n_2\right)^4} 
\nonumber\\
& \; \times \left(\frac{n-n_1}{n+n_1}\right)^{n+n_1} \,
\left(\frac{n-n_2}{n+n_2}\right)^{n+n_2} \times T_1 \times T_2 \,,
\end{align} 
where
\begin{align} 
T_1 =& \; {}_2F_1\left(2-n,1-n_1,2, u_1\right)
\nonumber\\
& \; -\frac{\left(n-n_1\right)^2}{\left(n+n_1\right)^2} \,
{}_2F_1\left(-n,1-n_1,2, u_1\right) \,,
\nonumber\\
T_2 =& \; {}_2F_1\left(2-n,1-n_2,2, u_2\right)
\nonumber\\
& \; -\frac{\left(n-n_2\right)^2}{\left(n+n_2\right)^2} \, 
{}_2F_1\left(-n,1-n_2,2, u_2\right) \,,
\end{align} 
and the arguments of the hypergeometric functions are
\begin{equation}
u_1 = -\frac{4 n\, n_1}{\left(n-n_1\right)^2} \,,
\qquad
u_2 = -\frac{4 n\, n_2}{\left(n-n_2\right)^2} \,.
\end{equation}
\end{subequations} 
As it stands, formula~\eqref{resG1}
is not applicable to the case $n = n'$
and has to be supplemented by the result
\begin{equation}
\label{resG2}
G_{nnn}^{010} = \frac{9}{4}\, n^2 \, (n^2-1)\, e^2\, a_0^2 \,.
\end{equation}
 
An alternative 
representation of the transition matrix elements,
to encompass both formulas~\eqref{resG1} and~\eqref{resG2},
would thus be desirable.
However, a literature search including Sec.~2.19.4 of~\cite{PrBrMa2002vol2}
does not reveal any immediately applicable integral formulas for 
integrals of the type~\eqref{4int}.
However, an entry in a recently published online 
database~\cite{FormulaLLE} allows us to express 
the integral~\eqref{4int} as a finite nested double sum
over terms involving the Pochhammer 
symbol $(a)_n = \Gamma(a + n)/\Gamma(a)$, 
\begin{multline}
\int_0^\infty \mathrm{d}t\  t^{\alpha-1}  
\exp\left(-p\, t\right) \ {L}_m^{\lambda}\left(a\,t\right) 
L_n^{\beta}\left(b\,t\right) 
\\
= \frac{\Gamma(\alpha) \left(\lambda+1\right)_m 
\left(\beta+1\right)_n p^{-\alpha}}{m! \, n!} \,
\\
\quad \times
\sum_{j=0}^{m}
\frac{ \left(-m\right)_j (\alpha)_j }{j! \left(\lambda+1\right)_j} \,
\left(\frac{a}{p}\right)^j
\sum_{k=0}^{n}\frac{ \left(-n\right)_k (j+\alpha)_k }{k!  \left(\beta+1\right)_k} 
\left(\frac{b}{p}\right)^k \,.
\end{multline} 
The inner sum can be expressed in terms of a
terminating hypergeometric function.
The coefficient $G^{010}_{n_1nn_2}$ is 
finally written in a rather compact form, as follows,
\begin{multline}
\label{totalintIIf1st}
G^{010}_{n_1nn_2} = 
e^2 \, a_0^2 \,  2^{10}\, \frac{\left(n_1 n_2\right)^{7/2}\, 
n^5\, \left(n^2-1\right)}{\left(n+n_1\right)^5\, \left(n+n_2\right)^5} 
\\
\times \sum_{\zeta=0}^{n_1-1}
\frac{ \left(1-n_1\right)_{\zeta} (5 )_{\zeta}}{\zeta!  \left(2\right)_{\zeta}} 
\left(\frac{2 n}{n_1+ n}\right)^\zeta 
\\
\times {}_2F_1\left(2-n, 5+\zeta,  4,  \frac{2 n_1}{n+n_1}\right) 
\\
\times \sum_{\beta=0}^{n_2-1}
\frac{ \left(1-n_2\right)_{\beta} (5 )_{\beta} }{\beta!  \left(2\right)_{\beta}} 
\left(\frac{2 n}{n_2+ n}\right)^\beta 
\\
\times {}_2F_1\left(2-n, 5+\beta,  4,  \frac{2 n_2}{n+n_2}\right) .
\end{multline}
The case $n_1 = n_2 = n$, which is excluded from the 
treatment described in Ref.~\cite{Go1929aop}, 
is important in the derivation of Eq.~\eqref{eq:D6BarPattern}.

\end{document}